\documentclass{aa}
\def\deg{\ifmmode^{\circ}\;\else$^{\circ}\;$\fi} 
\def\lsim{\,\lower2truept\hbox{${<\atop\hbox{\raise4truept\hbox{$\sim$}}}$}\,}
\def\gsim{\,\lower2truept\hbox{${> \atop\hbox{\raise4truept\hbox{$\sim$}}}$}\,}
\usepackage{graphicx}
\begin{document}

\thesaurus{06}    

   \title{A multifrequency analysis of radio variability of blazars}
   \subtitle{ }

   \author{
        A. Ciaramella\inst{1,7}
        \and
          C. Bongardo\inst{2}
        \and
        H.D. Aller\inst{3}
        \and
        M.F. Aller\inst{3}
        \and
        G. De Zotti \inst{2}
        \and
        A. L\"{a}hteenmaki \inst{4}
        \and
        G. Longo \inst{5,6}
        \and
        L. Milano \inst{5,6}
        \and
        R. Tagliaferri
        \inst{1,7}
       \and
          H. Ter\"asranta\inst{4}
          \and
          M. Tornikoski\inst{4}
          \and
          S. Urpo\inst{4}
          }

   \offprints{A. Ciaramella: ciaram@unisa.it}

\institute{
Dipartimento di Matematica ed Informatica, Universit\`a
di Salerno, via S. Allende, Baronissi (Sa), Italy\\
   \and
   INAF--Osservatorio Astronomico di Padova, Vicolo dell'Osservatorio 5, I-35122 Padova,
   Italy\\
   \and Dept.\ of Astronomy, Dennison Bldg., U.\ Michigan, Ann Arbor, MI
   48109, USA \\
   \and Mets\"ahovi Radio Observatory, 02540 Kylm\"al\"a, Finland \\
   \and
   Dipartimento di Scienze Fisiche, Universit\`a Federico II, via Cinthia 6, I-80126 Napoli, Italy
   \and
   INFN - Sezione di Napoli, via Cinthia 6, I-80126 Napoli, Italy
   \and
   INFM - Sezione di Salerno, via S. Allende, Baronissi (SA), Italy
   }

   \date{Received ..... ; accepted ........}

 \maketitle

\begin{abstract}
We have carried out a multifrequency analysis of the radio
variability of blazars, exploiting the data obtained during the
extensive monitoring programs carried out at the University of
Michigan Radio Astronomy Observatory (UMRAO, at 4.8, 8, and 14.5
GHz) and at the Mets\"ahovi Radio Observatory (22 and 37 GHz). Two
different techniques detect, in the Mets\"ahovi light curves,
evidences of periodicity at both frequencies for 5 sources
($0224+671$, $0945+408$, $1226+023$, $2200+420$, and $2251+158$).
For the last three sources consistent periods are found also at
the three UMRAO frequencies and the Scargle (1982) method yields
an extremely low false-alarm probability. On the other hand, the
22 and 37 GHz periodicities of $0224+671$ and $0945+408$ (which
were less extensively monitored at Mets\"ahovi and for which we
get a significant false-alarm probability) are not confirmed by
the UMRAO database, where some indications of ill-defined periods
about a factor of two longer are retrieved. We have also
investigated the variability index, the structure function, and
the distribution of intensity variations of the most extensively
monitored sources. We find a statistically significant difference
in the distribution of the variability index for BL Lac objects
compared to flat-spectrum radio quasars (FSRQs), in the sense that
the former objects are more variable. For both populations the
variability index steadily increases with increasing frequency.
The distribution of intensity variations also broadens with
increasing frequency, and approaches a log-normal shape at the
highest frequencies. We find that variability enhances by 20--30\%
the high frequency counts of extragalactic radio-sources at bright
flux densities, such as those of the WMAP and {\sc Planck}
surveys. In all objects with detected periodicity we find evidence
for the existence of impulsive signals superimposed on the
periodic component.

     \end{abstract}


%

\section{Introduction}

The name Blazars identifies a family of radio-loud Active Galactic
Nuclei (AGNs) showing a rather complex phenomenology: extreme
variability at all wavelengths, polarization, strong $\gamma$--ray
emission, brightness temperatures exceeding the Compton limit
(\cite{Urry1999}). The large amount of work done in the last
decade has led to a rather general consensus on the global
mechanism responsible for the emission: a rotating black hole
surrounded by a massive accretion disk with an intense plasma jet
closely aligned to the line of sight. Relativistic electrons
produce the soft photons through synchrotron emission, while hard
photons are produced by inverse Compton scattering. This overall
scenario, however, still presents a large number of poorly
understood details which, in turn, lead to a wide variety of
models and call for long term and multi--wavelength campaigns
capable to provide the necessary observational constraints.

Variability measurements provide key information on the AGN
structure, down to linear scales or flux density levels not
accessible with interferometric imaging. The most extensive blazar
monitoring campaigns have been carried out at radio frequencies.
The University of Michigan monitoring program at 4.8, 8.0 and 14.5
GHz has obtained data on over 200 sources for over three decades.
At higher frequencies, the Mets\"ahovi group (Ter\"asranta et al.
1998) have reported observations at 22, 37 and 87 GHz of 157
extragalactic radio sources, many of which have been monitored for
over 20 years. The Bologna group (Bondi et al. 1996) have observed
at 408 MHz 125 radio sources for 15 years. A multifrequency
monitoring of large sample of compact radio sources has been
carried out by Kovalev et al. (2002) with the RATAN-600 telescope.

The mechanisms for variability are still not well understood.
Possibilities discussed in the literature include shocks in jets
(Marscher \& Gear 1985; Aller, Aller \& Hughes 1985; Marscher
1996) and changes in the direction of forward beaming [due, e.g.
to helical trajectories of plasma elements (Camenzind \&
Krockenberger 1992) or to a precessing binary black-hole system
(Begelman et al. 1980; Sillanp\"a\"a et al. 1988)], introducing
flares due to the lighthouse effect. Thus, variability furnishes
important clues into size, structure, physics and dynamics of the
radiating source region.

The cm-wavelength variability of extragalactic sources has been
investigated by Hughes et al. (1992) and Aller et al. (1996, 2003)
exploiting the uniquely large, long term, University of Michigan
Radio Astronomy Observatory (UMRAO) data-base. Kelly et al. (2003)
applied a cross-wavelet transform to analyze the UMRAO light
curves of the Pearson-Readhead VLBI survey sources. The
Mets\"ahovi data-base has been analyzed by Lainela \& Valtaoja
(1993), Ter\"asranta \& Valtaoja (1994), Valtaoja et al. (1999),
L\"ahteenm\"aki et al. (1999), L\"ahteenm\"aki \& Valtaoja (1999).

Still, the information content of the radio-monitoring data-base
has not yet been fully exploited. We are interested, in
particular, in utilizing it to predict the effect of radio-source
variability on the high-frequency all-sky surveys to be carried
out by ESA's {\sc Planck} mission. Such predictions are useful to
plan the quick-look analysis of {\sc Planck} data and to organize
follow-up observations. To this end, the Mets\"ahovi data, taken
at frequencies close to those of {\sc Planck}-LFI (Low Frequency
Instrument) channels, are particularly well fit.

Another very interesting, highly debated issue, is the possible
presence of periodicities in the blazar light curves. Claims for
the existence of periodic behavior are mostly based on optical
data (Lainela et al. 1999; Fan 2000, Fan et al. 2002, and
references therein). The most famous, but still somewhat
controversial, case is the 11.65 year periodicity of OJ287
(Sillanp\"a\"a et al. 1988, 1996; Marchenko et al. 1996;
Hagen-Thorn et al. 1997; Pietil\"a et al. 1999). Evidence of a
persistent modulation of the radio total flux and polarization of
this source, with period of $\sim 1.66\,$yr was found by Aller et
al. (1992) and Hughes et al. (1998).

A possible periodicity of $\sim 5.7$ years in the radio
light-curve of the BL Lac object AO$\,0235+16$ has been reported
by Roy et al. (2000) and Raiteri et al. (2001). Indications of a
periodic or quasi-periodic component with a time-scale of 5.5--6
years (close to that of AO$\,0235+16$) of the radio emission of
the BL Lac object S5$\,0716+714$ were found by Raiteri et al.
(2003)

In this paper we present a new investigation of radio light curves
of a sample of blazars, exploiting more effective techniques than
the Periodogram method commonly used (in various versions) for
periodicity analysis. The maximum time interval covered by the
data we used is 1979--2001 in the case of Mets\"ahovi, 1965--1999
in the case of UMRAO. The techniques used are described in Section
2 and the main results are presented in Section 3. In Section 4 we
discuss the multifrequency variability properties for a blazar
sample monitored by both the UMRAO and the Mets\"ahovi Radio
Observatory, and estimate the effect of variability on
high-frequency counts of such objects. In Section 5 we summarize
our main conclusions.












\begin{table*}
\caption[]{Objects for which we have positive detection of
periodicity on Mets\"ahovi daily averaged data. Column 1: object
identification. Columns 2-5: 22 GHz data; columns 6-9: 37 GHz
data. Columns 2 \& 6: number of data points; columns 3 \& 7:
maximum admissible period to avoid aliasing; columns 4 \& 8:
period (in units of $10^3$ days) obtained by STIMA; column 5 \& 9:
period obtained from the Lomb's Periodogram.} \label{perioMetsa_1}
\centering
\begin{tabular}{l|rrrr|rrrr|l}
 \hline
 \hline
  Name    & N & Mx. P. & STIMA     & Lomb & N & Mx. P. & STIMA     & Lomb    &  Notes\\
          &         & ($\times 10^3$)&($\times 10^3$)&($\times
          10^3$)&
          & ($\times 10^3$)&($\times 10^3$)&($\times 10^3$)&    \\
\multicolumn{1}{c|}{(1)} & \multicolumn{1}{c}{(2)} &
\multicolumn{1}{c}{(3)} & \multicolumn{1}{c}{(4)} &
\multicolumn{1}{c|}{(5)} & \multicolumn{1}{c}{(6)} &
\multicolumn{1}{c}{(7)} & \multicolumn{1}{c}{(8)} &
\multicolumn{1}{c|}{(9)} & \multicolumn{1}{c}{(10)} \\
 \hline
$0224+671$   & 76      & 3.062& 1.021& 1.021& 63  &3.668&0.950&0.970&      \\
$0945+408$   & 47      & 3.742& 1.386& 1.336& 26
&4.467&1.313&1.240& few
points    \\
$ 1226+023$   & 694     & 6.665& 3.029& 2.777& 716 &7.822&3.260&1.261&      \\
$2200+420$   & 644     & 6.676& 3.034& 3.034& 715 &7.873&2.811&3.028&      \\
$2251+158$   & 571     & 6.676& 2.384& 2.384& 549 &7.538&2.217&2.512&      \\
\end{tabular}
\end{table*}

%


\begin{table*}

\caption[]{Analysis of UMRAO data on Mets\"ahovi sources with
detected periodicity (Table~\ref{perioMetsa_1}). Columns 2-5, 6-9,
and 10-13 refer to 4.8, 8.0, and 14.5 GHz data, respectively. }
\label{perioUMRAO}

 \centering
\begin{tabular}{l|rrrr|rrrr|rrrr|}
 \hline
 \hline
  Name    &  N & Mx. P. & STIMA     & Lomb  & N &Mx. P. & STIMA     & Lomb & N &Mx. P. & STIMA     & Lomb \\
          &         & ($\times 10^3$)&($\times 10^3$)&($\times 10^3$)
          &         & ($\times 10^3$)&($\times 10^3$)&($\times 10^3$)
          &         &($\times 10^3$)&($\times 10^3$)&($\times 10^3$)\\
\multicolumn{1}{c|}{(1)} & \multicolumn{1}{c}{(2)} &
\multicolumn{1}{c}{(3)} & \multicolumn{1}{c}{(4)} &
\multicolumn{1}{c|}{(5)} & \multicolumn{1}{c}{(6)} &
\multicolumn{1}{c}{(7)} & \multicolumn{1}{c}{(8)} &
\multicolumn{1}{c|}{(9)} & \multicolumn{1}{c}{(10)} &
\multicolumn{1}{c}{(11)} & \multicolumn{1}{c}{(12)} &
\multicolumn{1}{c|}{(13)} \\
 \hline
$0224+671$ & 110 & 5.998 & 1.764 & 1.764 & 119 & 6.859 & 2.286 &
2.286 & 110 & 7.592 & 2.109 & 2.109 \\
$0945+408$ & 73 & 6.048
&\multicolumn{1}{c}{--}& \multicolumn{1}{c|}{--} &
    126& 7.993&  \multicolumn{1}{c}{--}    & \multicolumn{1}{c|}{--}      &   92& 5.920 &  2.690&  2.466\\
$1226+023$ & 493  & 6.424 &  3.197& 3.197& 1294& 1.232& 3.082& 3.082 &  760& 9.114 &  3.038&  3.038\\
$2200+420$ & 745  & 7.844 &  2.801& 2.801& 1212& 1.131& 2.261& 1.413 & 1056& 9.056 &  2.830&  2.830\\
$2251+158$ & 554  & 7.674 &  2.240& 2.240& 1175& 1.195& 2.390& 2.490 &  885& 9.190 &  2.297&  2.297\\
\end{tabular}

\end{table*}

\begin{table*}
\caption[]{Summary of detected periods (in units of $10^3$ days).
} \label{periosummary}

\centering
\begin{tabular}{lrrrrr}
 \hline
 \hline
  Name    & 4.8 GHz & 8.0 GHz & 14.5 GHz& 22 GHz& 37 GHz\\
 \hline
 0224+671   &  1.764 & {2.286}   & {2.109}  & 1.021 & {0.950}\\
 0945+408   &  \multicolumn{1}{c}{--} & \multicolumn{1}{c}{--}  & 2.690
 & {1.386} & {1.313} \\
 1226+023   & 3.197 & 3.082  & 3.038 & 3.029 & {3.260}\\
 2200+420   & 2.801 & 2.261  & 2.830 & 3.034 & {2.811}\\
 2251+158   & 2.240 & 2.390  & 2.297 & 2.384 & {2.217}\\
\end{tabular}
\end{table*}
\medskip

\begin{table}
{\small\caption{The U sample with source classifications by Donato
et al. (2001) and Ter\"{a}sranta  et al. (1998). In the latter
case BLO stands for BL Lac Object. Redshifts where taken from the
SIMBAD astronomical database. \label{Usample}} \centering
\begin{tabular}{lllll} \hline
Name & Other name & Donato & Ter\"{a}sranta  &
\multicolumn{1}{c}{$z$}
\\ \hline
0048-097 & OB -080 & LBL  & BLO  & 0.2    \\
0133+476$^\dag$ & OC 457  & FSRQ & HPQ & 0.859  \\
0202+149$^\dag$ & 4C 15.05& FSRQ & HPQ & 0.405  \\
0235+164$^\dag$ & OD 160  & LBL  & BLO  & 0.94   \\
0420-014$^\dag$ & OF -135 & FSRQ & HPQ & 0.915  \\
0430+052$^\dag$ & 3C 120  & FSRQ & LPQ$^*$ & 0.0331 \\
0605-085 & OH -010 &      & HPQ & 0.872  \\
0607-157$^\ddag$ & OH 112  &      &      & 0.324 \\
0735+178$^\dag$ & OI 158  & LBL  & BLO  & 0.424  \\
0814+425$^\dag$ & OJ 425  & LBL  & BLO  & 0.2453 \\
0851+202$^\dag$ & OJ 287  & LBL  & BLO  & 0.306  \\
0923+392$^\dag$ & 4C 39.25  & FSRQ & LPQ & 0.6948 \\
0945+408 & 4C 40.24& FSRQ & LPQ & 1.252  \\
0954+658 &         & LBL  & BLO  & 0.367  \\
1034-293 & OL -259 & FSRQ & HPQ$^*$ & 0.312  \\
1055+018$^\dag$ & OL 093  & FSRQ & HPQ & 0.888  \\
1127-145$^\ddag$ & OM -146 &      &      & 1.187 \\
1226+023$^\dag$ & 3C 273  & FSRQ & LPQ & 0.158  \\
1253-055$^\dag$ & 3C 279  & FSRQ & HPQ & 0.538  \\
1308+326$^\dag$ & OP 313  & LBL  & BLO  & 0.996  \\
1418+546$^\dag$ & OQ 530  & LBL  & BLO  & 0.151  \\
1510-089$^\dag$ & OR -017 & FSRQ & HPQ & 0.359  \\
1538+149 & OR 165  & LBL  & BLO  & 0.605  \\
1637+574$^\dag$ & OS 562  &      & LPQ & 0.7506 \\
1641+399$^\dag$ & 3C345   & FSRQ & HPQ & 0.594  \\
1642+690 & 4C 69.21&      & HPQ & 0.751  \\
1749+096$^\dag$ & OT 081  & LBL  & BLO  & 0.322  \\
1749+701 &         & LBL  & BLO  & 0.77   \\
1823+568 & 4C 56.27& LBL  & BLO  & 0.6635 \\
1928+738 & 4C 73.18& FSRQ & LPQ & 0.360  \\
2131-021 & 4C -02.18& LBL & BLO  & 1.285  \\
2134+004$^\dag$ & OX 057  & FSRQ & LPQ & 1.931997\\
2145+067$^\dag$ & OX 076  &      & LPQ & 0.99   \\
2155-152$^\ddag$ & OX -192 & FSRQ &      & 0.672 \\
2200+420$^\dag$ & BL Lac  & LBL  & BLO  & 0.0688 \\
2201+315$^\dag$ & 4C 31.63&      & LPQ & 0.298  \\
2223-052$^\dag$ & 3C 446  & FSRQ & HPQ & 1.404  \\
2251+158$^\dag$ & 3C 454.3& FSRQ & HPQ & 0.859  \\
2254+074$^\dag$ & OY 091  & LBL  & BLO  & 0.19   \\ \hline
\ \\
\multicolumn{5}{l}{{}$^\dag$ {Also in the Mets\"{a}hovi database.}} \\
\multicolumn{5}{l}{{}$^\ddag$ {Classified as FSRQ by Stickel et al. (1994).}}\\
\multicolumn{5}{l}{{}$^*$ {Classified by Ghisellini et al. (1993).}}\\
\end{tabular} }
\end{table}

\begin{figure*}
\begin{center}
\includegraphics[height=4cm, width=6cm]{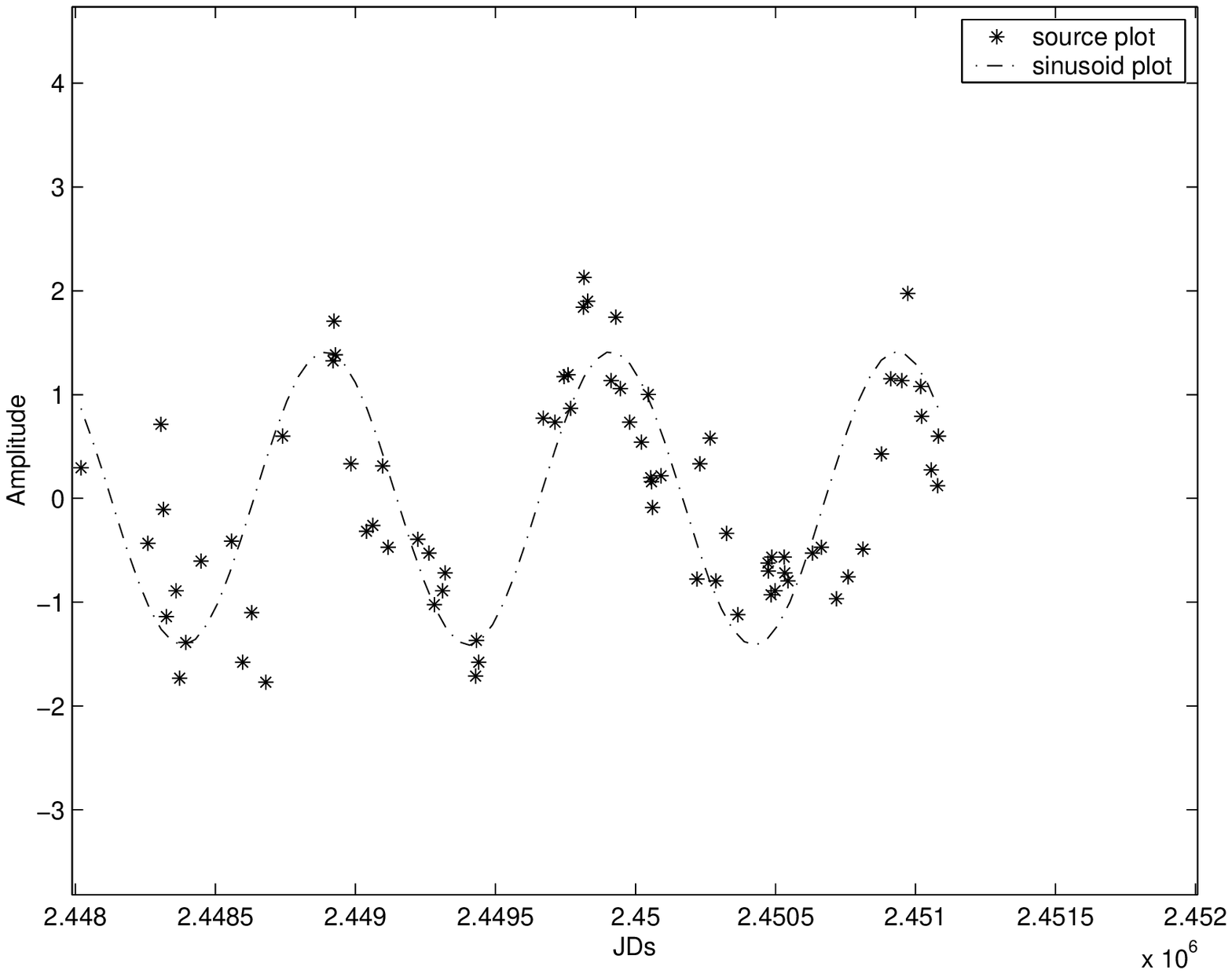}
\includegraphics[height=4cm, width=6cm]{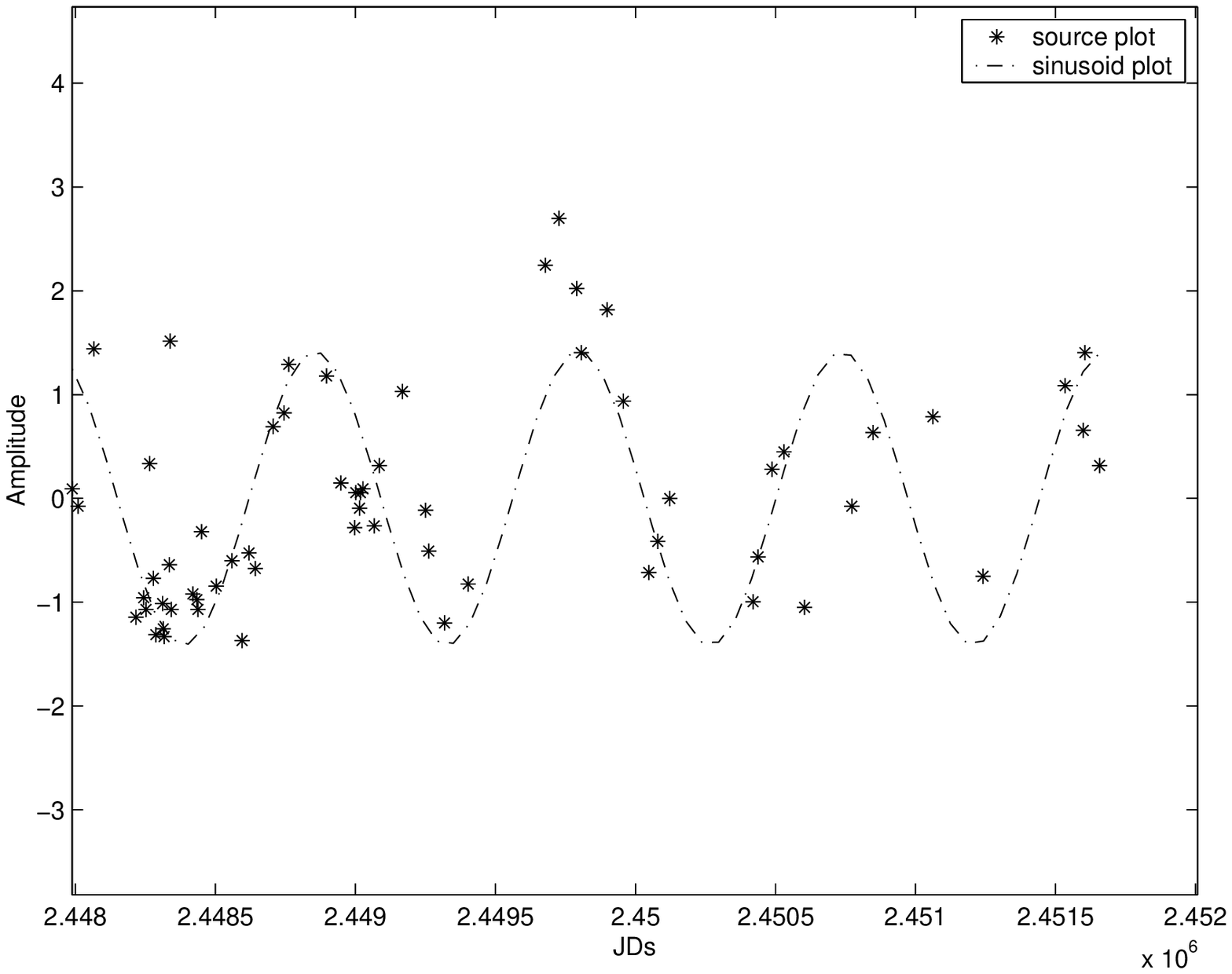}
\caption{A0224+671. Results for the $22$ GHz (left panel) and $37$
(right panel), daily averaged datasets. Sinusoids with a period
equal to those provided by STIMA and listed in Tables 1 and 2 are
overplotted as a visual aid to recognize periodicities. It needs
to be stressed that these curves must be regarded as visual aids
only: their amplitude is not by any means related to the
signal.}\label{A0224_1}
\end{center}
\end{figure*}

\begin{figure*}
\begin{center}
\includegraphics[height=4cm, width=6cm]{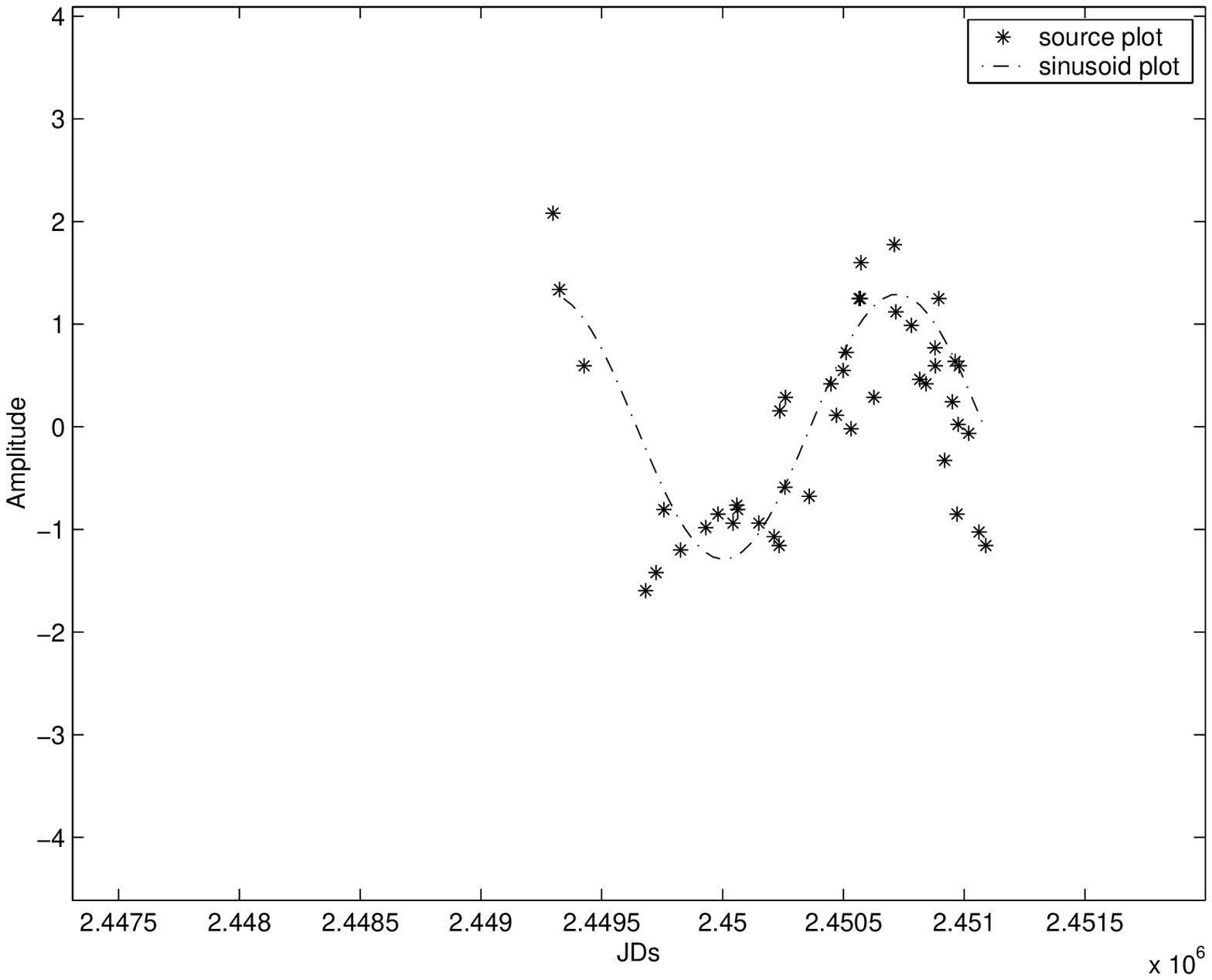}
\includegraphics[height=4cm, width=6cm]{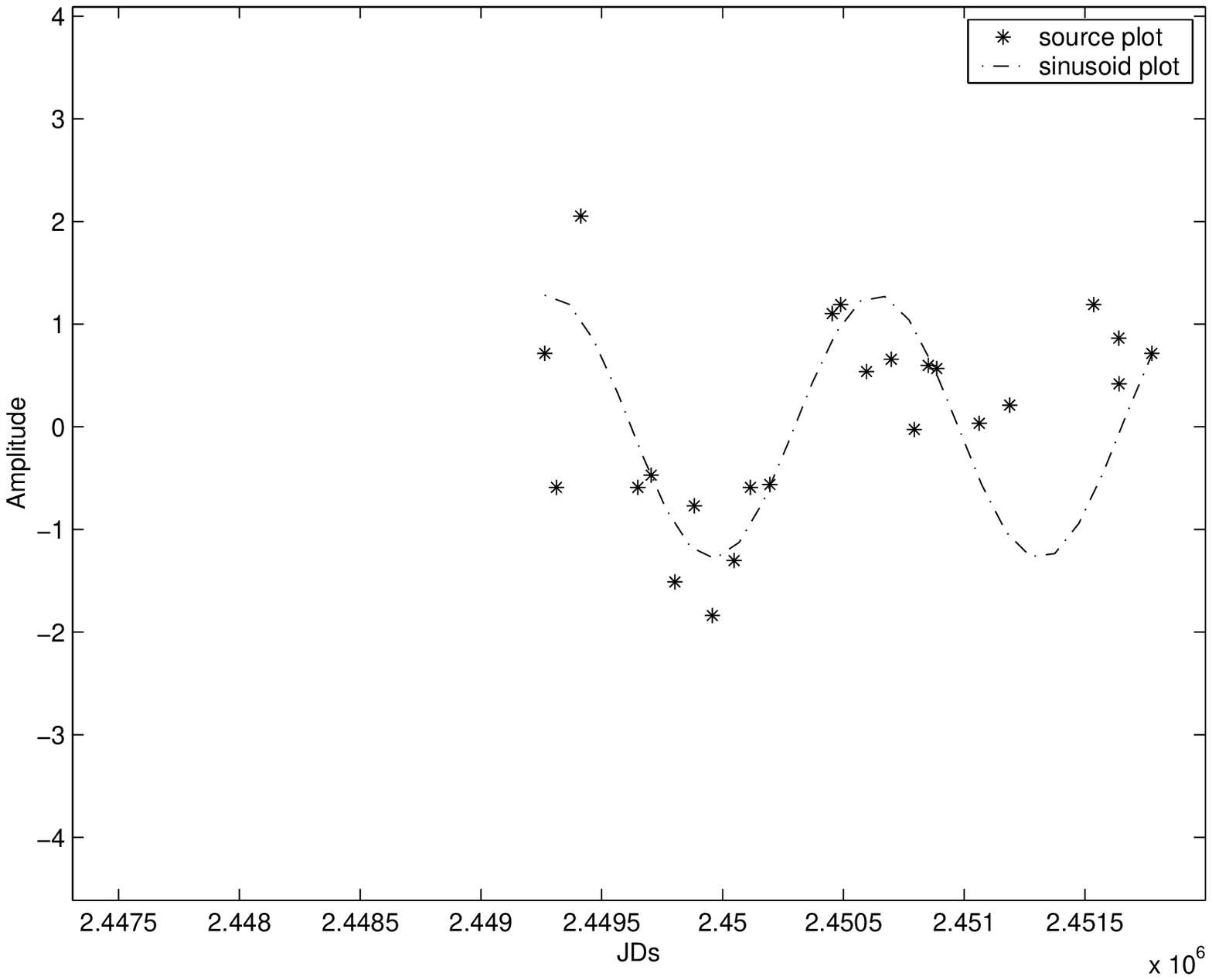})
\caption{$0945+408$. Results for the $22$ (left) and $37$ GHz
(right) $M$ dataset. Sinusoids as in Fig.
\ref{A0224_1}}\label{A094+540}
\end{center}
\end{figure*}

\begin{figure*}
\begin{center}
\includegraphics[height=4cm, width=6cm]{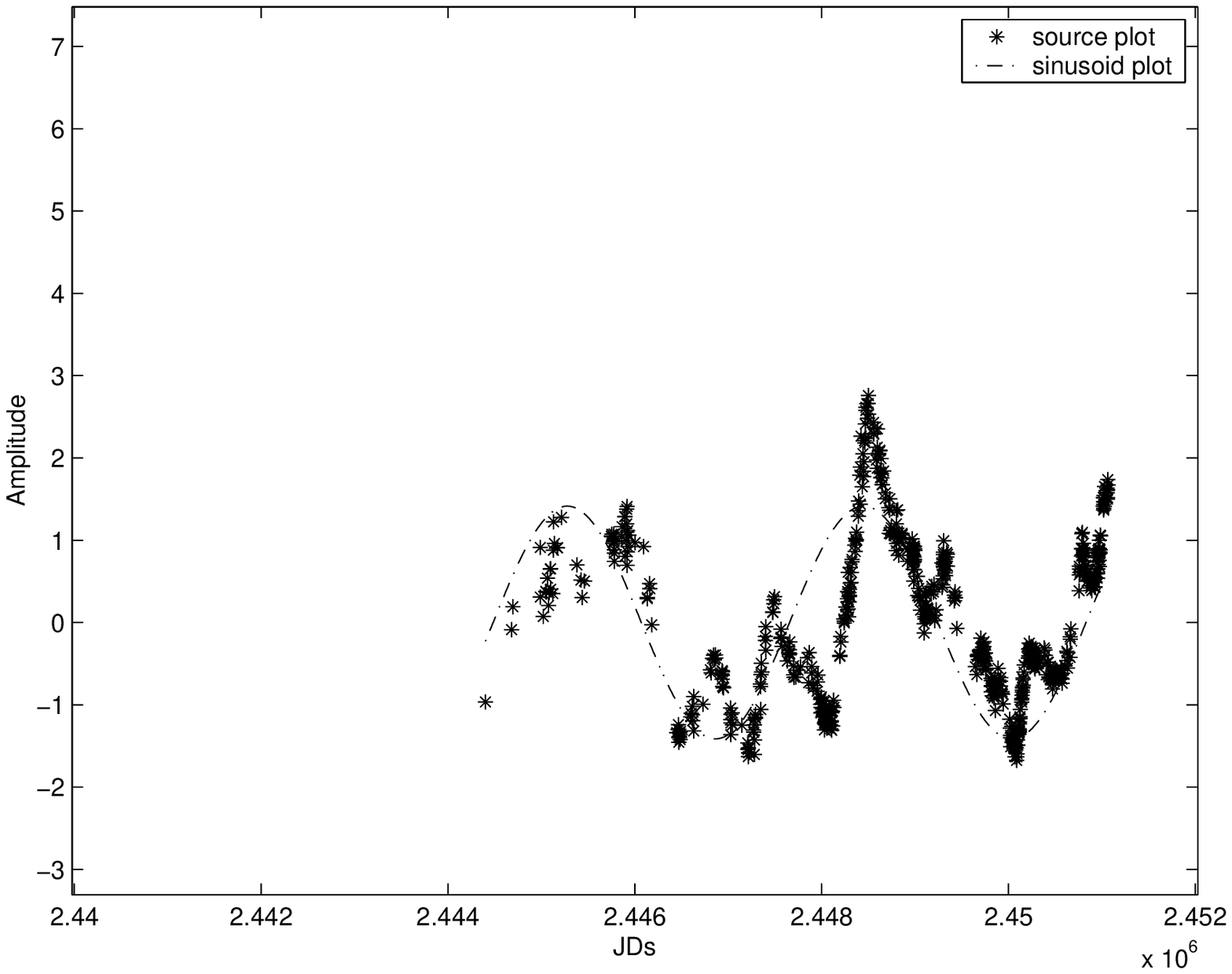}a)
\includegraphics[height=4cm, width=6cm]{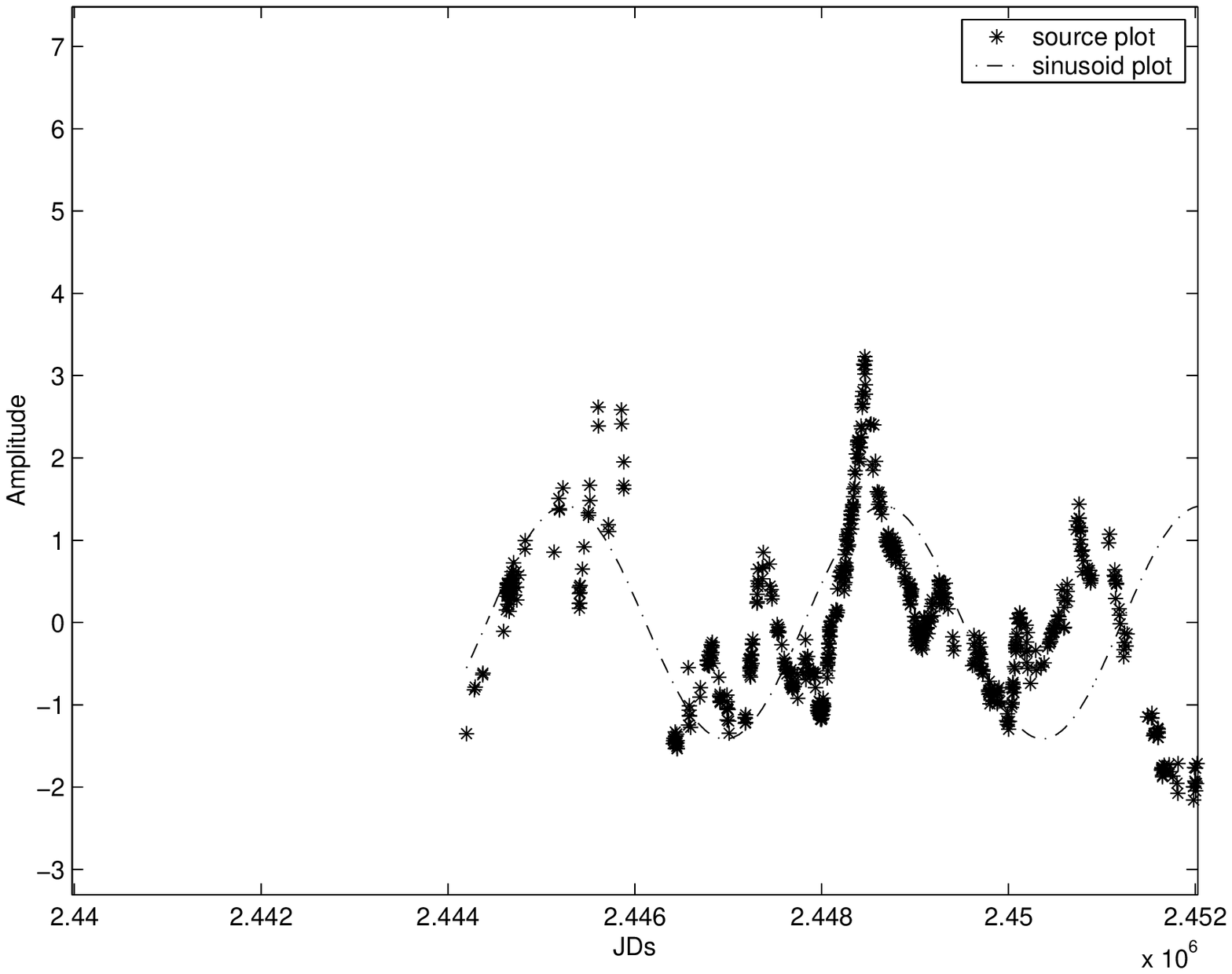}b)
\includegraphics[height=4cm, width=6cm]{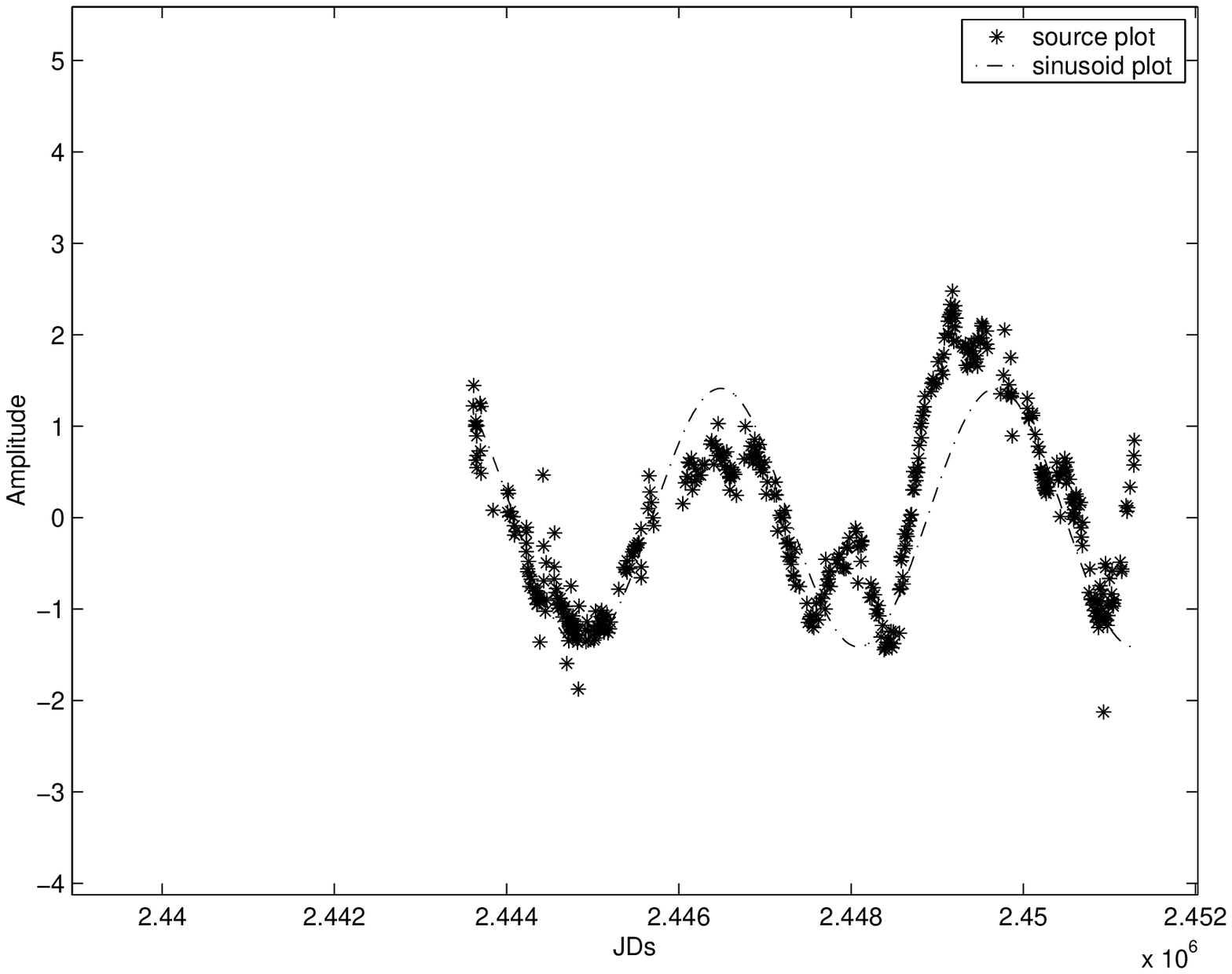}c)
\includegraphics[height=4cm, width=6cm]{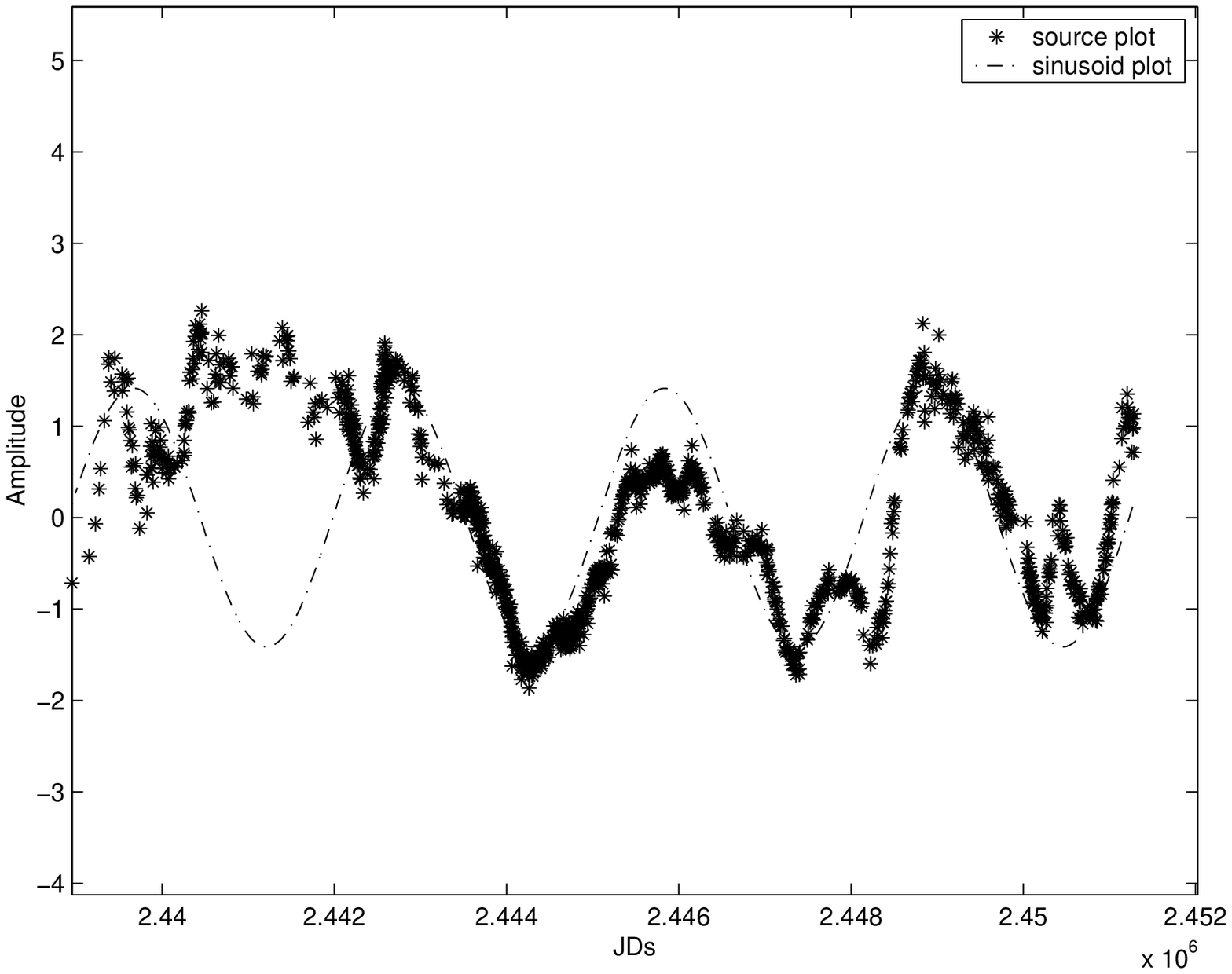}d)
\includegraphics[height=4cm, width=6cm]{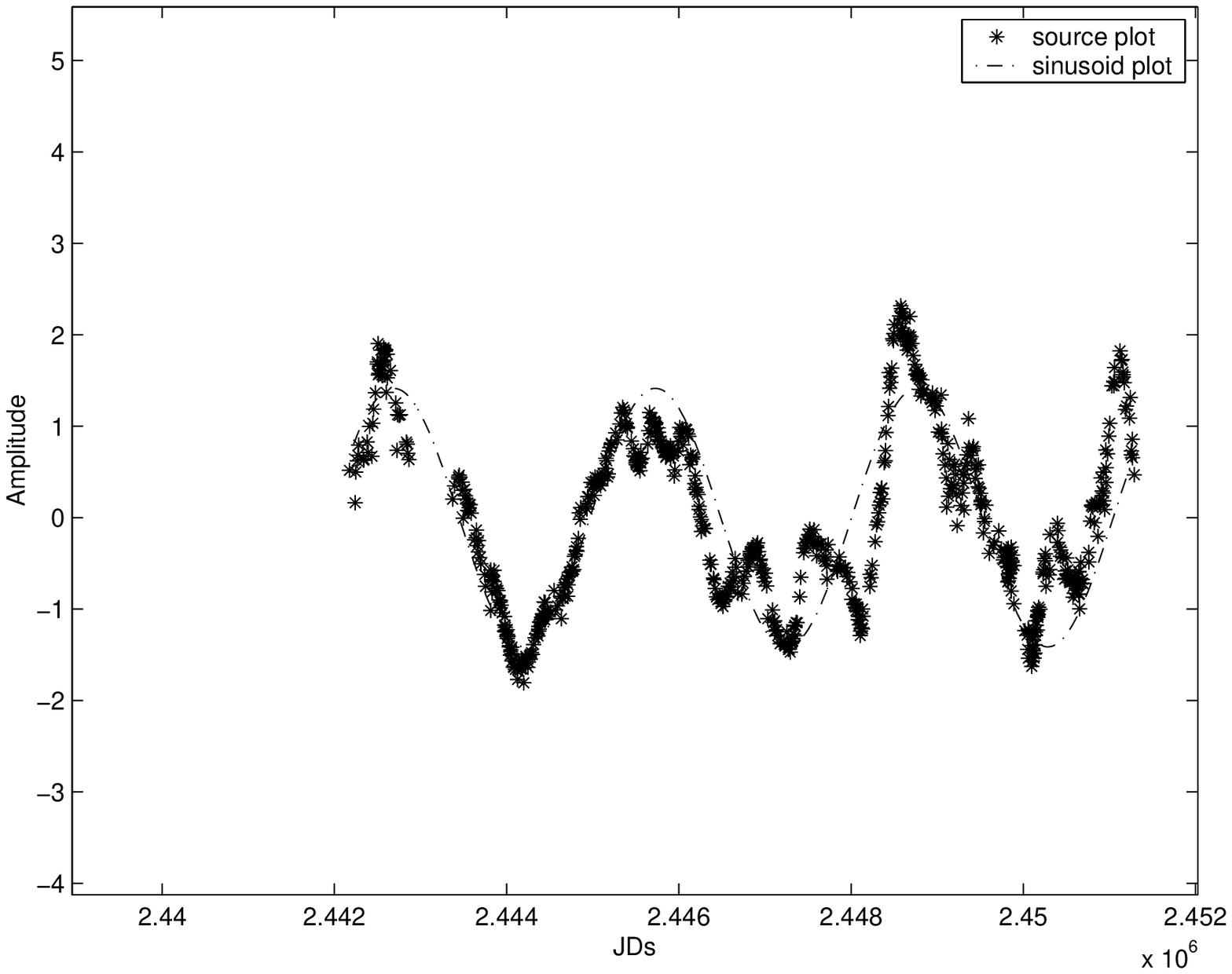}e)
\caption{1226+023. Panels a and b: results for the $22$ GHz and
the $37$ GHz daily averaged radio curves, respectively. Panels c,
d and e: the same for the $4.8$, $8$ and $14.5$ GHz data,
respectively. Sinusoids with a period equal to those listed in the
tables are overplotted as visual aid (as in
Fig.1).}\label{A1226_1}
\end{center}
\end{figure*}

\begin{figure*}
\begin{center}
\includegraphics[height=4cm, width=6cm]{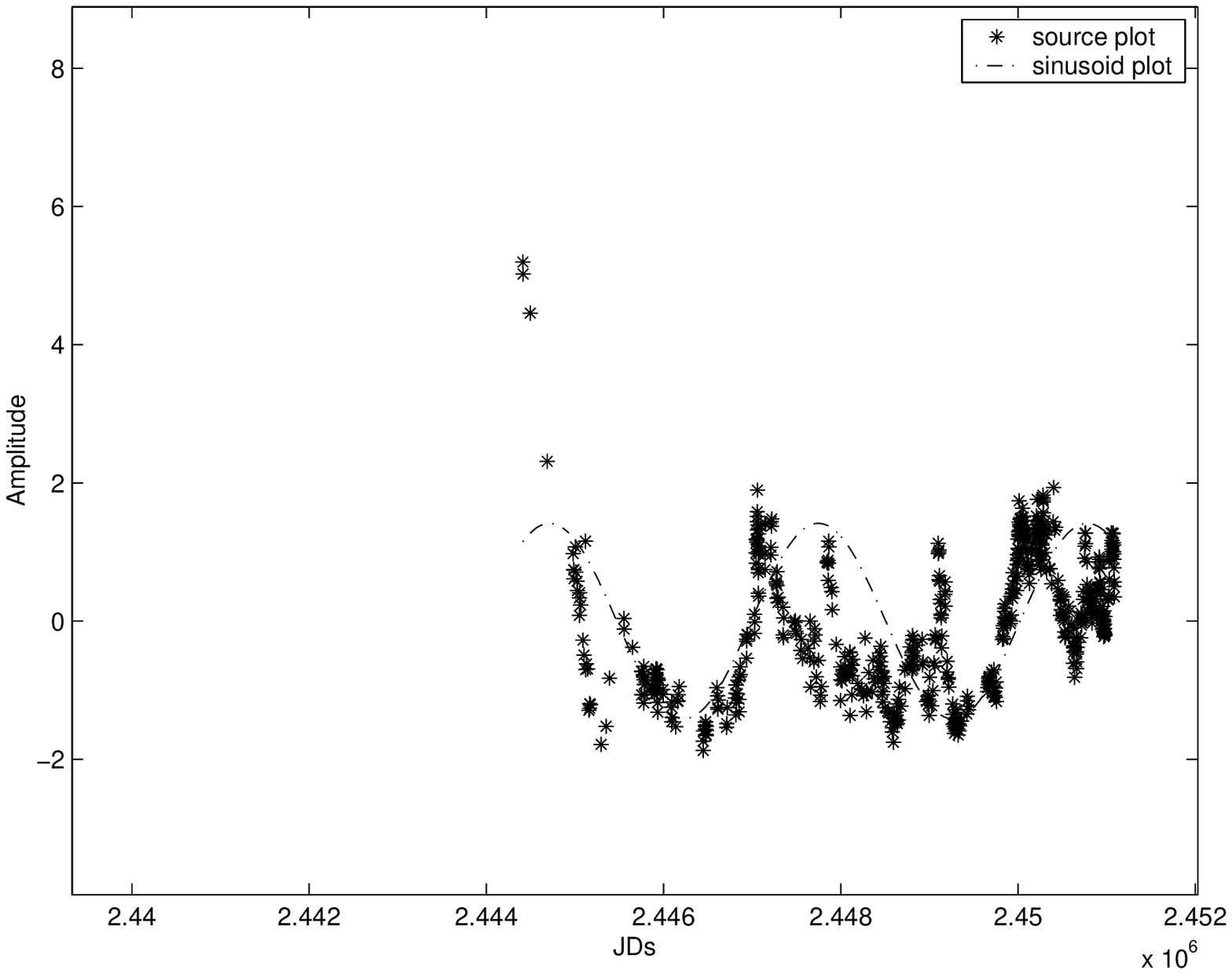}a)
\includegraphics[height=4cm, width=6cm]{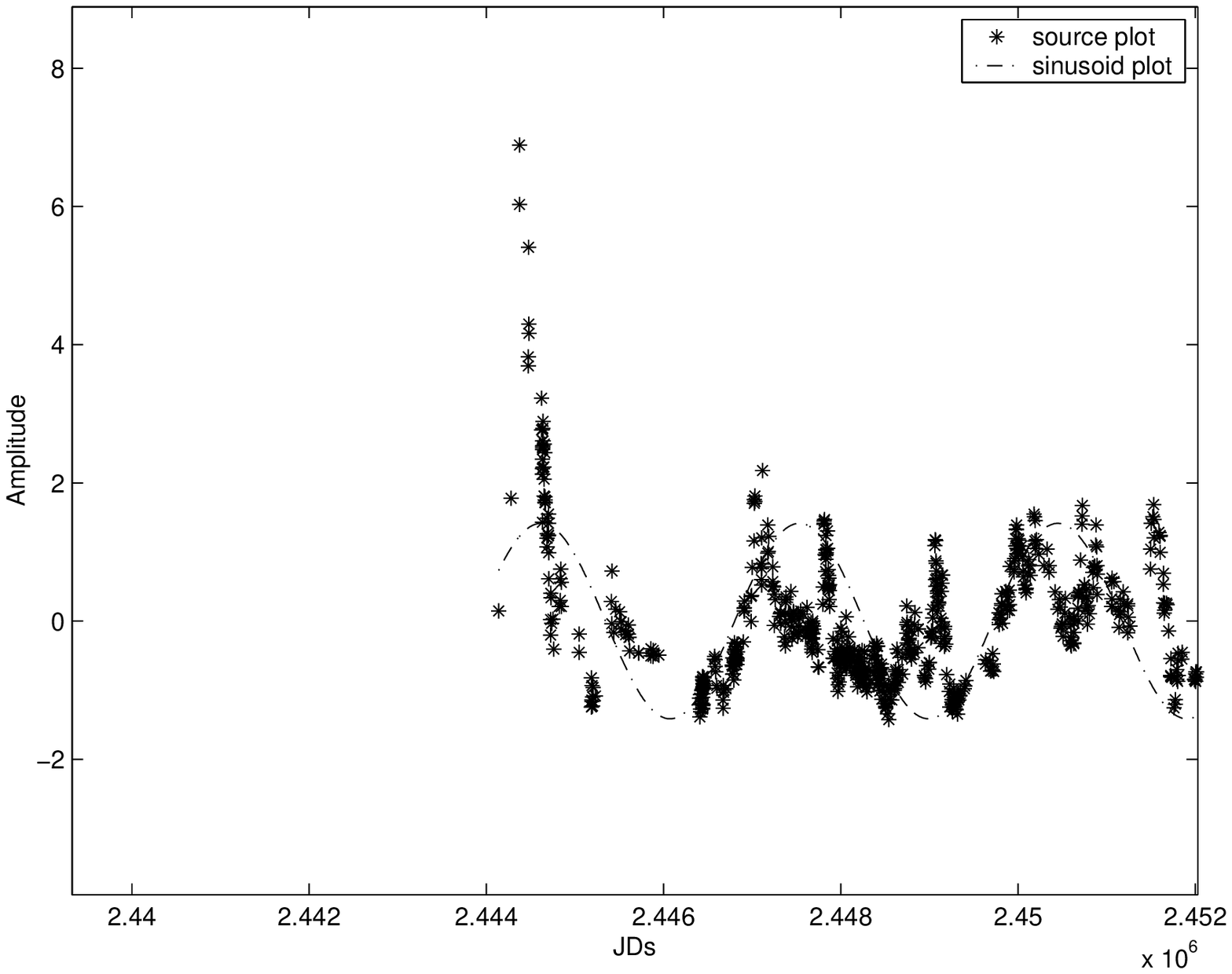}b)
\includegraphics[height=4cm, width=6cm]{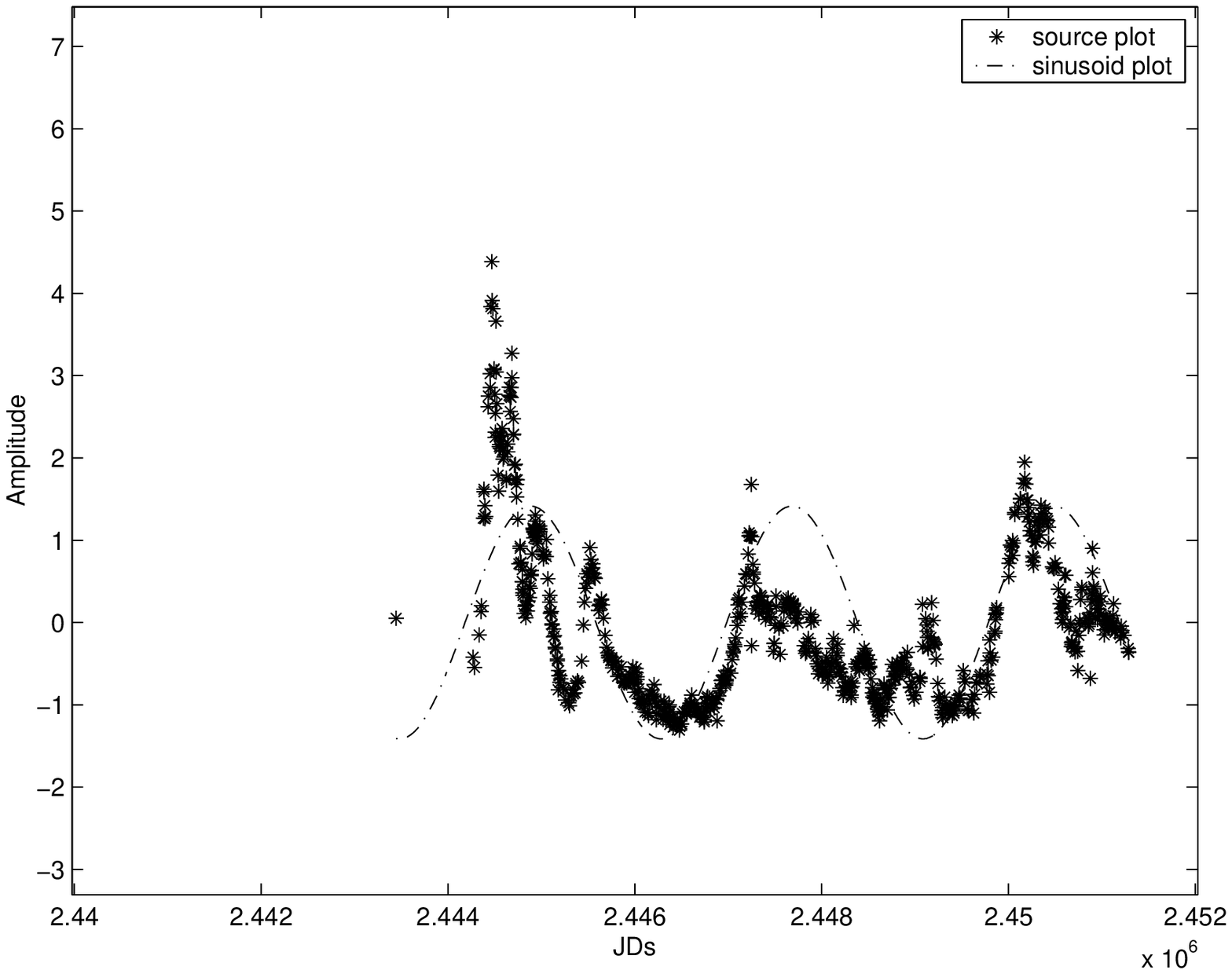}c)
\includegraphics[height=4cm, width=6cm]{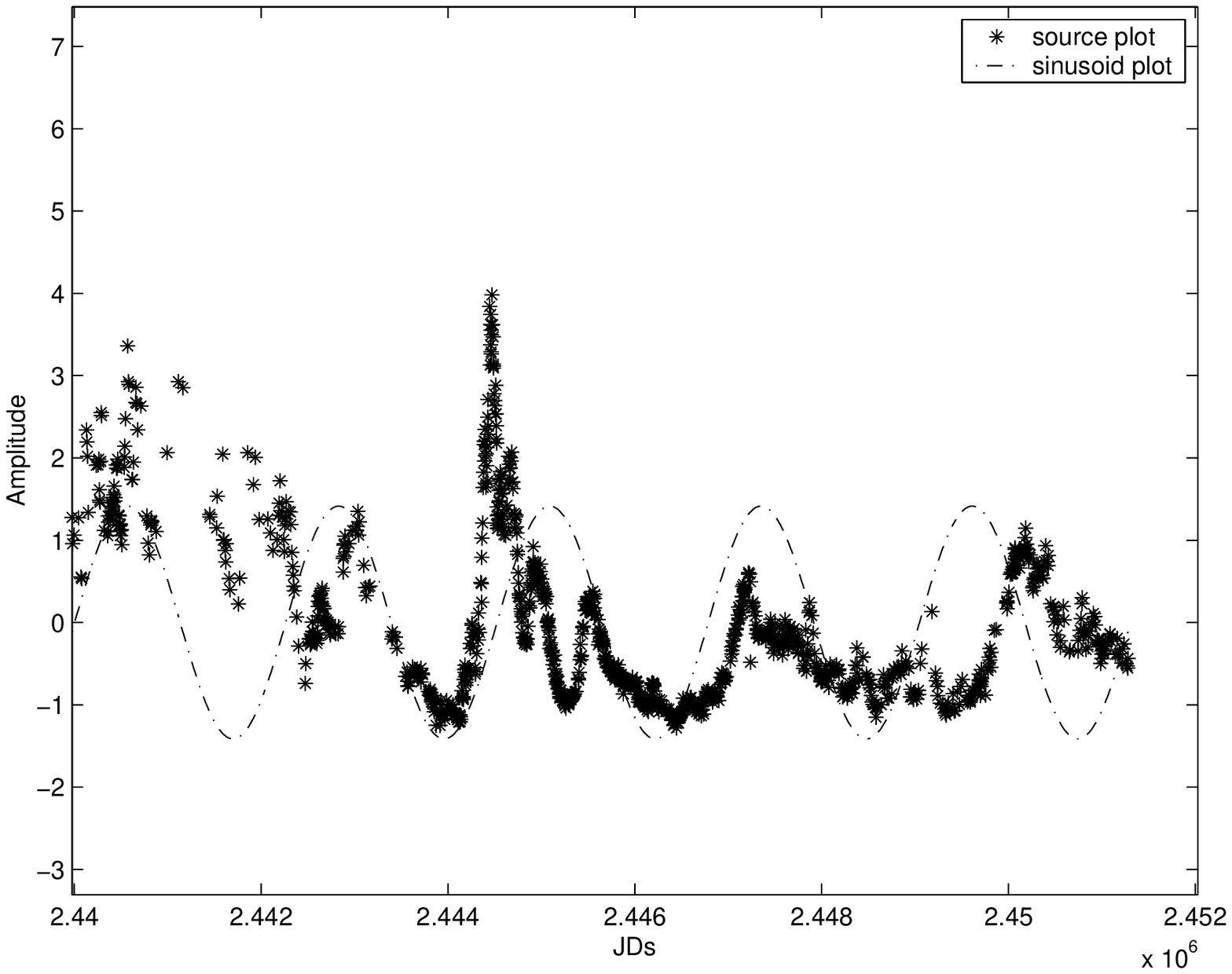}d)
\includegraphics[height=4cm, width=6cm]{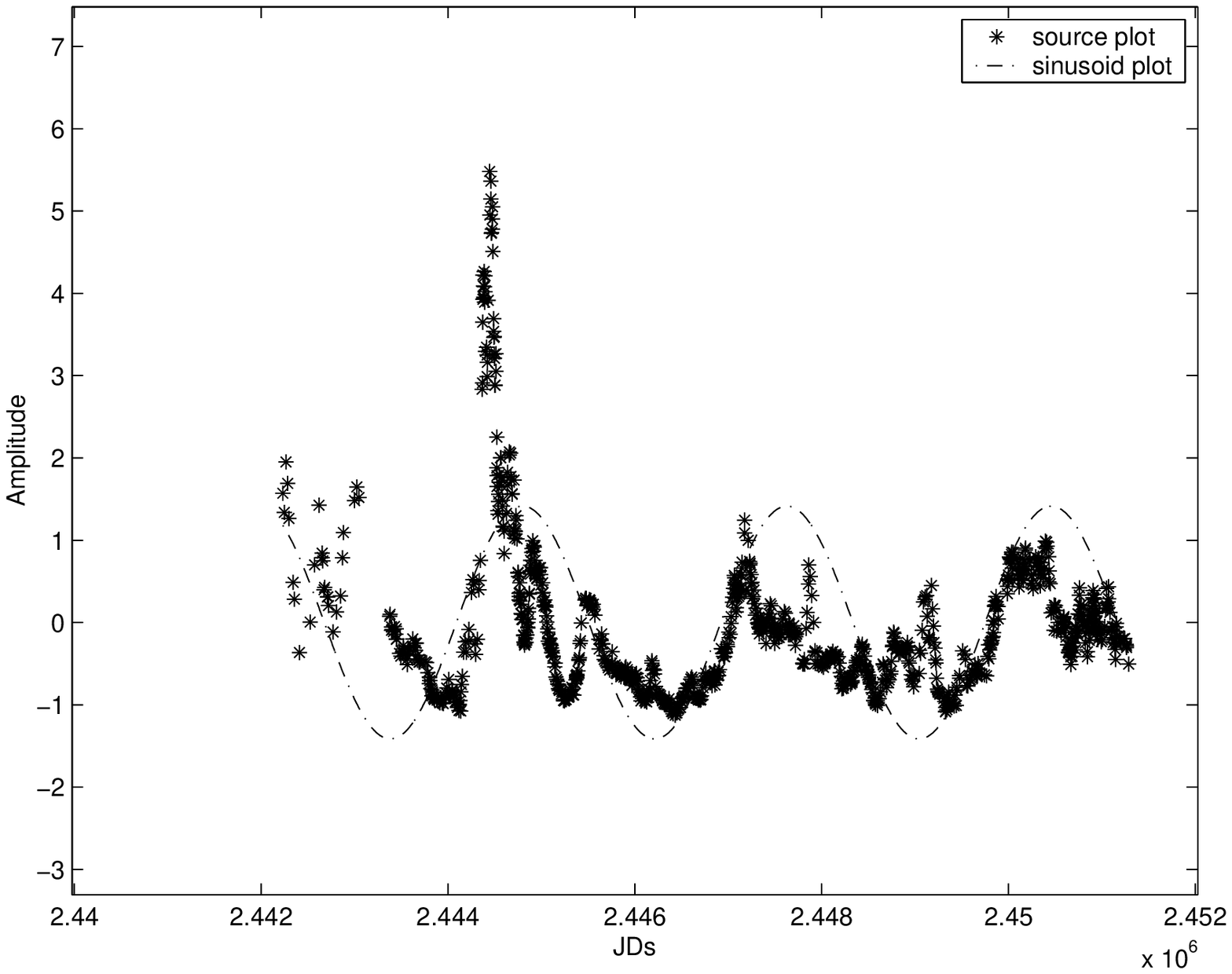}e)
\caption{$2200+420$. Panels have the same meaning as in
Fig.~\ref{A1226_1}.}\label{A2200_1}
\end{center}
\end{figure*}

\begin{figure*}
\begin{center}
\includegraphics[height=4cm, width=6cm]{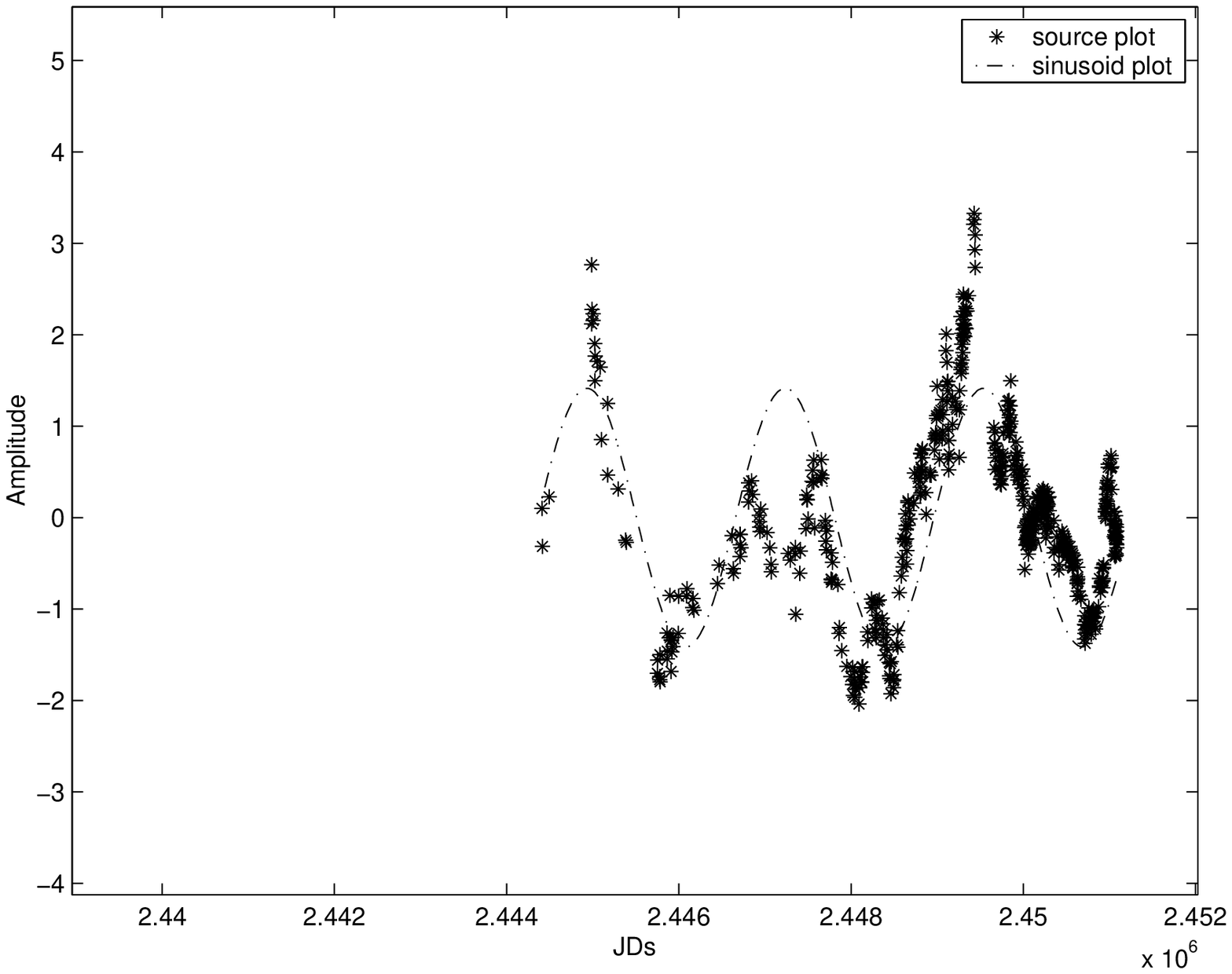}a)
\includegraphics[height=4cm, width=6cm]{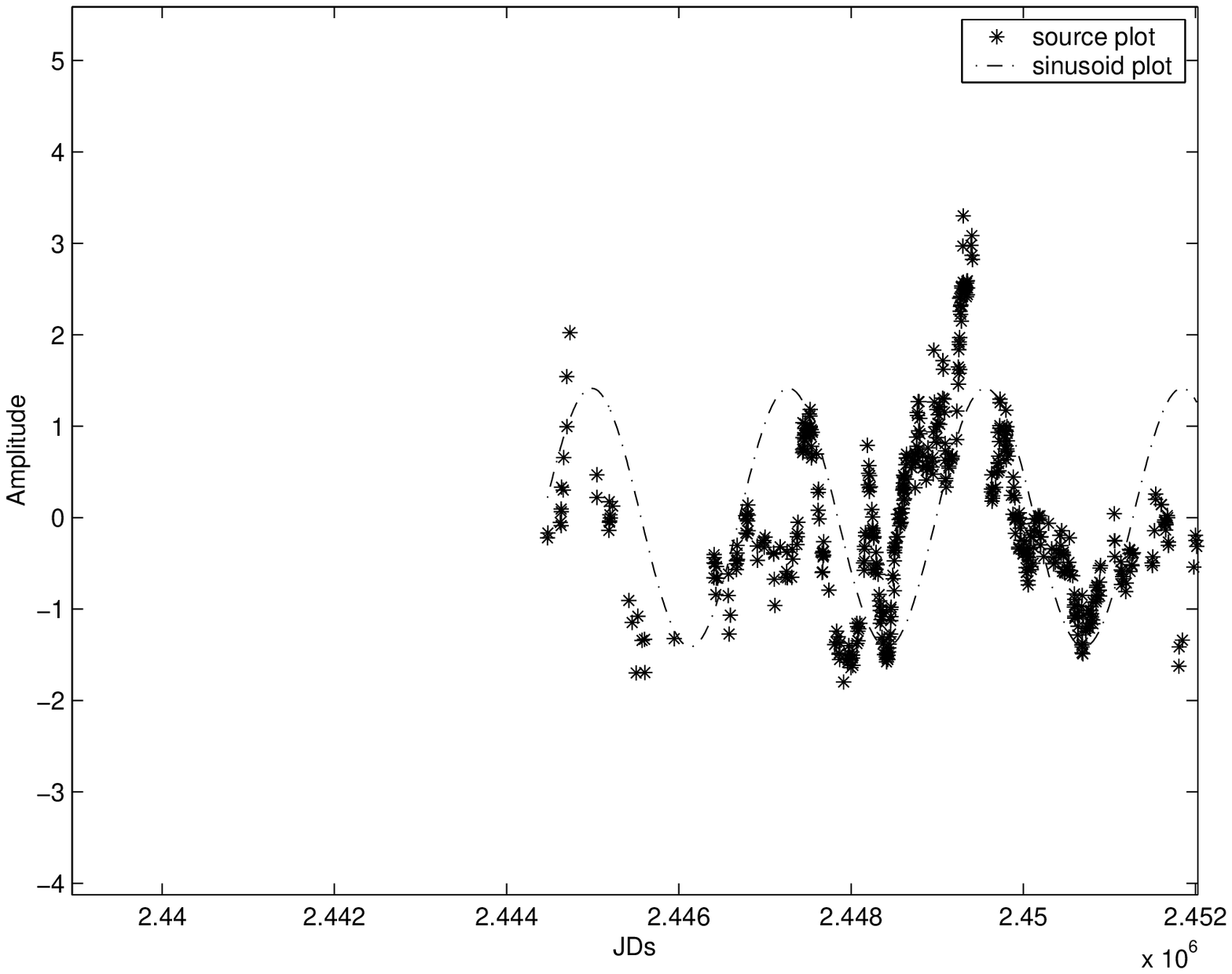}b)
\includegraphics[height=4cm, width=6cm]{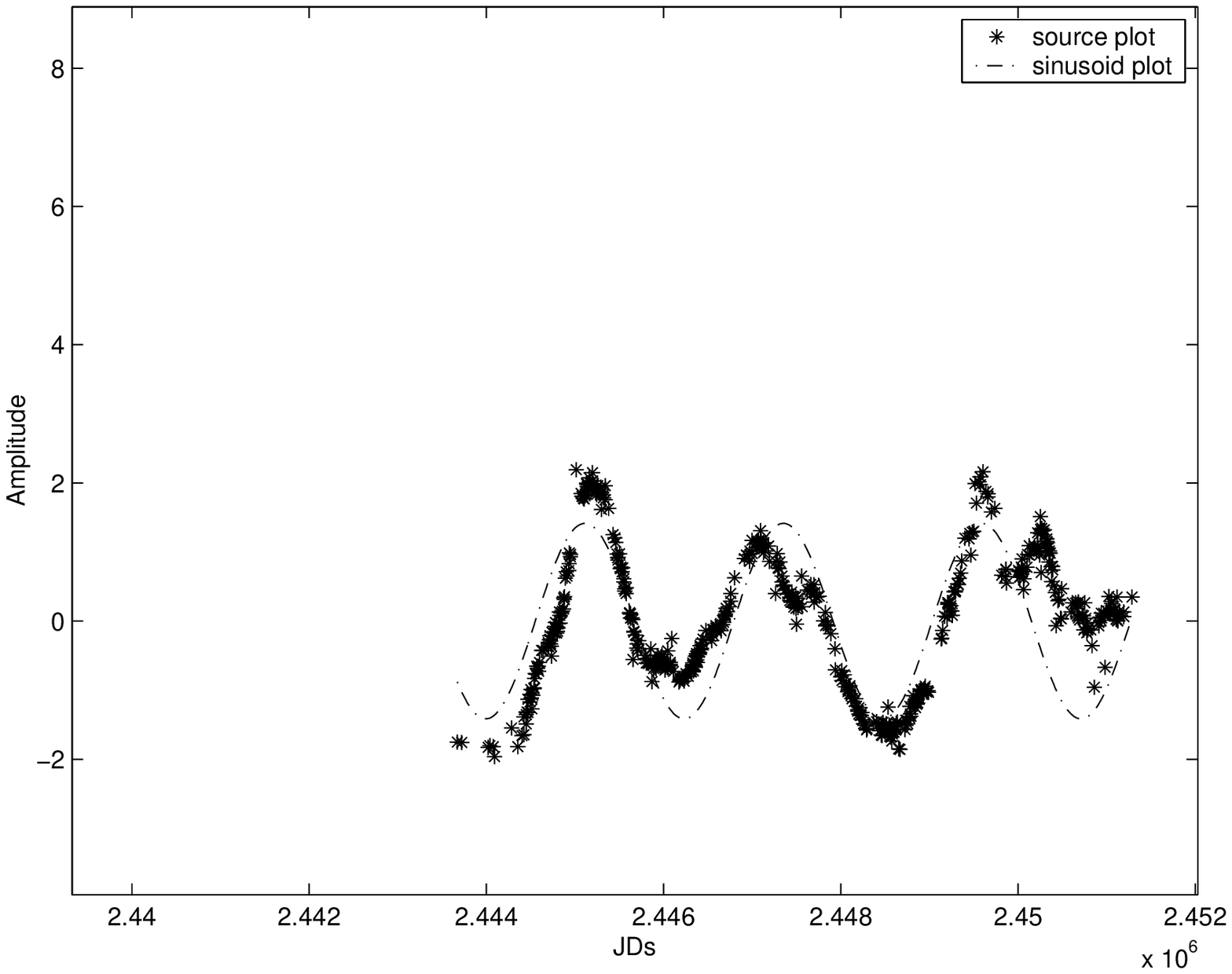}c)
\includegraphics[height=4cm, width=6cm]{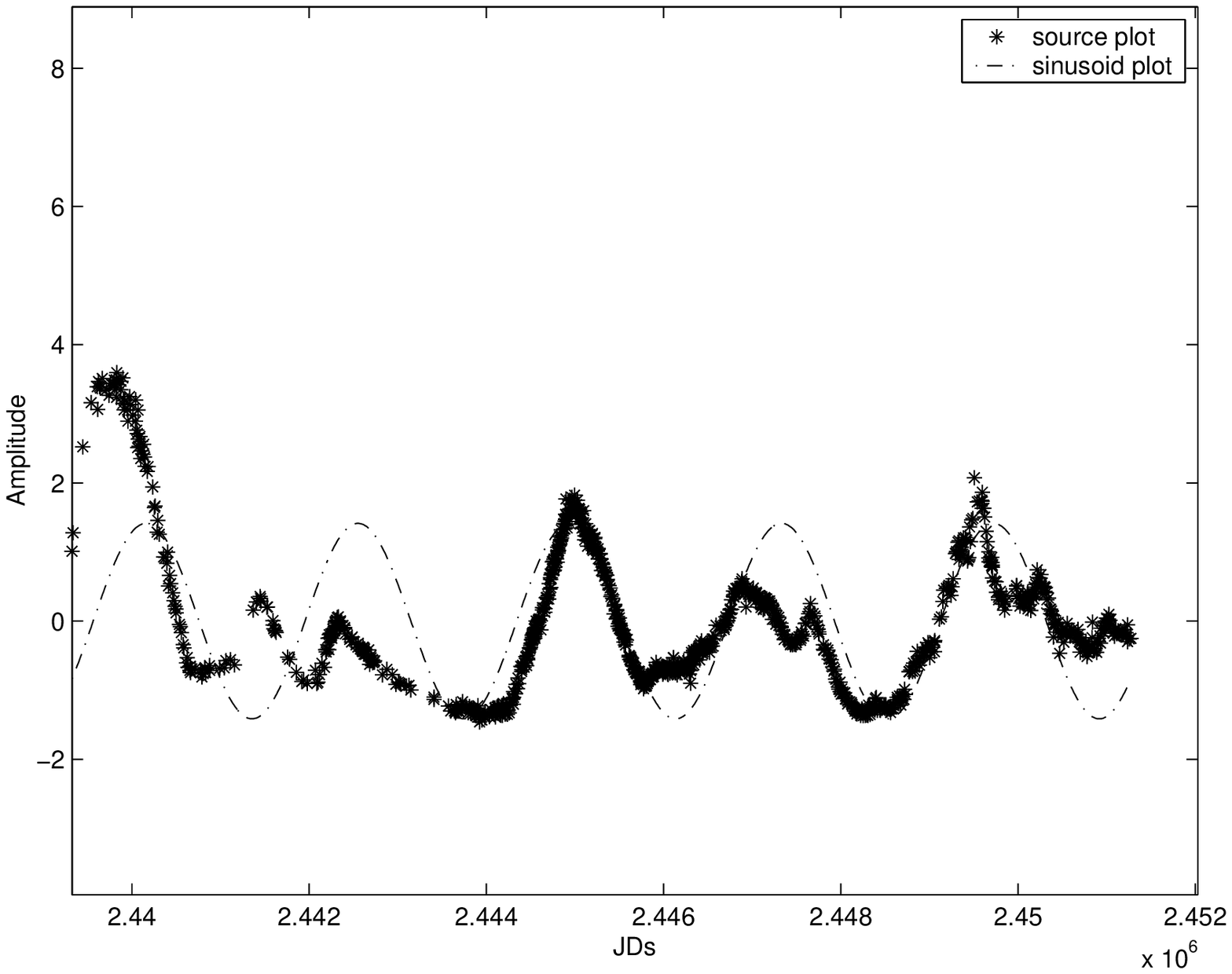}d)
\includegraphics[height=4cm, width=6cm]{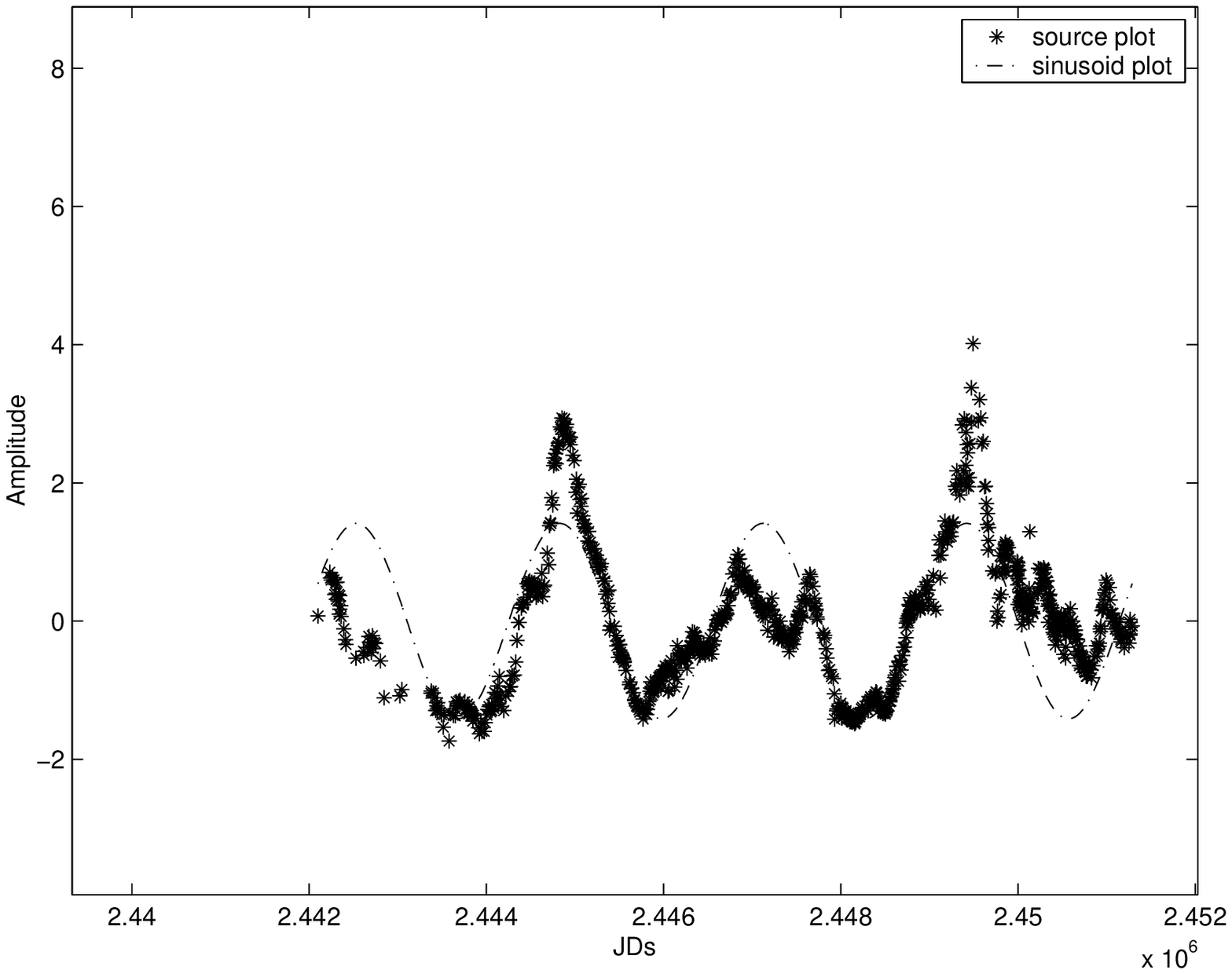}e)
\caption{$2251+158$. Panels have the same meaning as in
Fig.~\ref{A1226_1}.}\label{A2251_1}
\end{center}
\end{figure*}

\begin{figure*}
\begin{center}
\includegraphics[height=4cm, width=6cm]{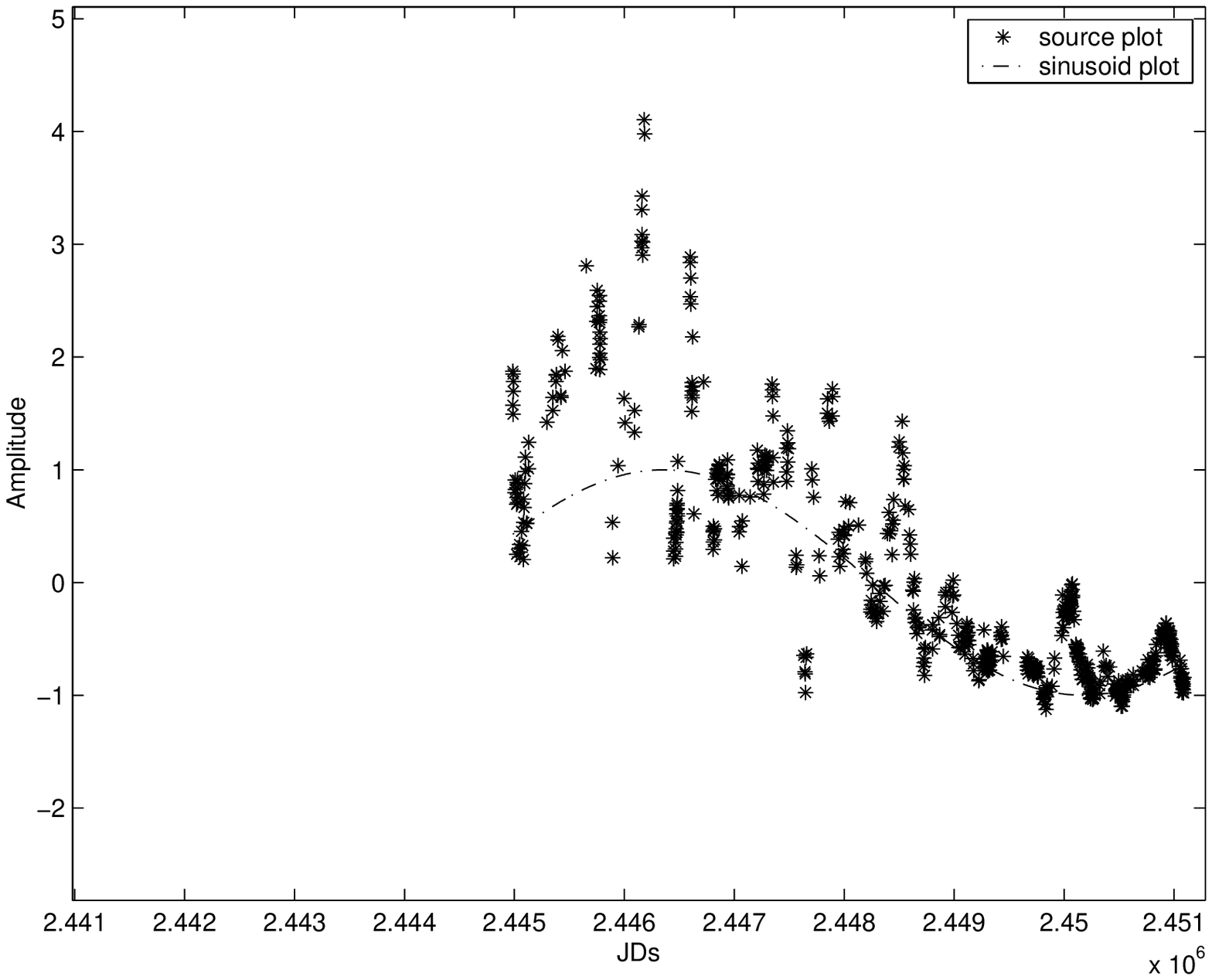}a)
\includegraphics[height=4cm, width=6cm]{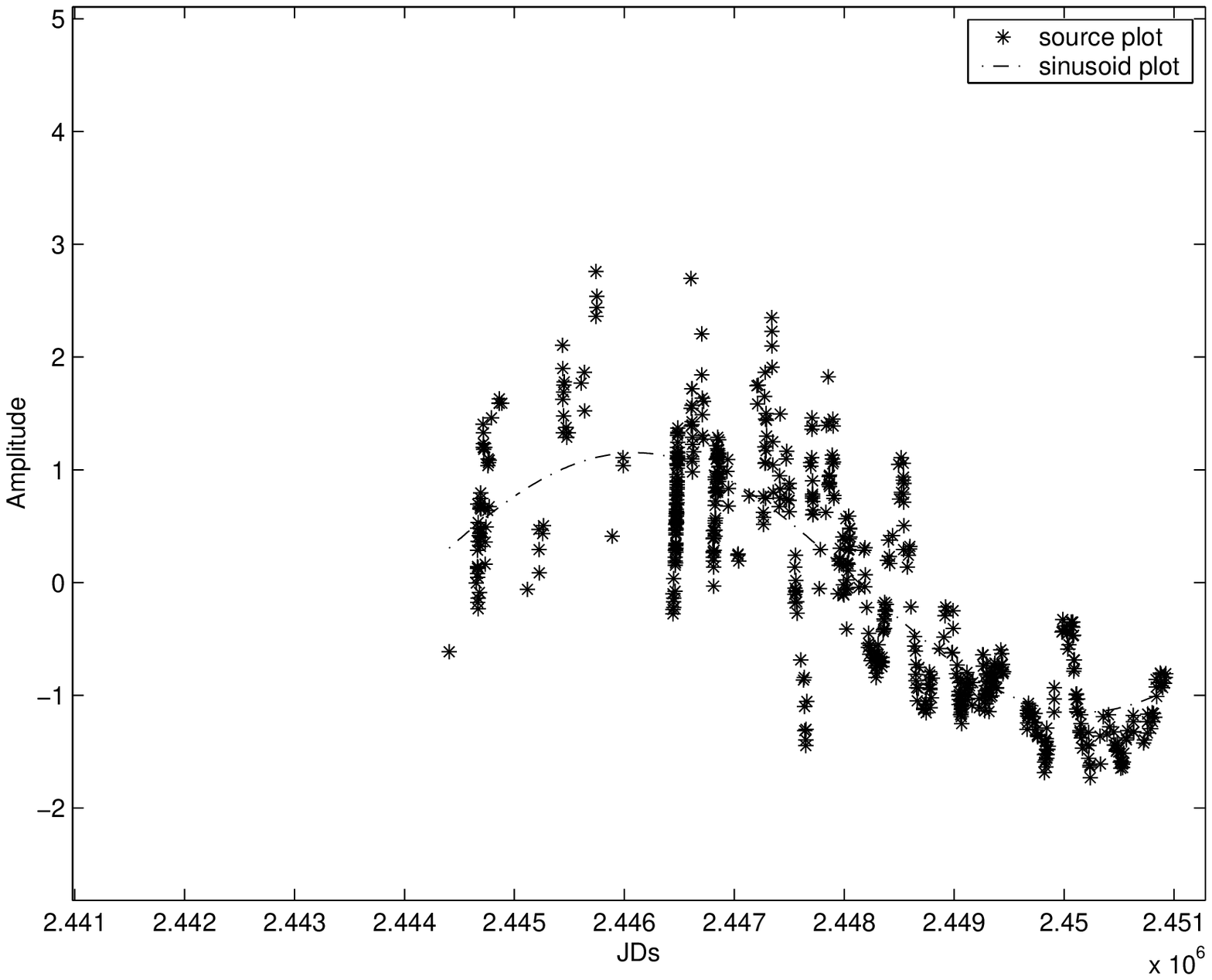}b)
\includegraphics[height=4cm, width=6cm]{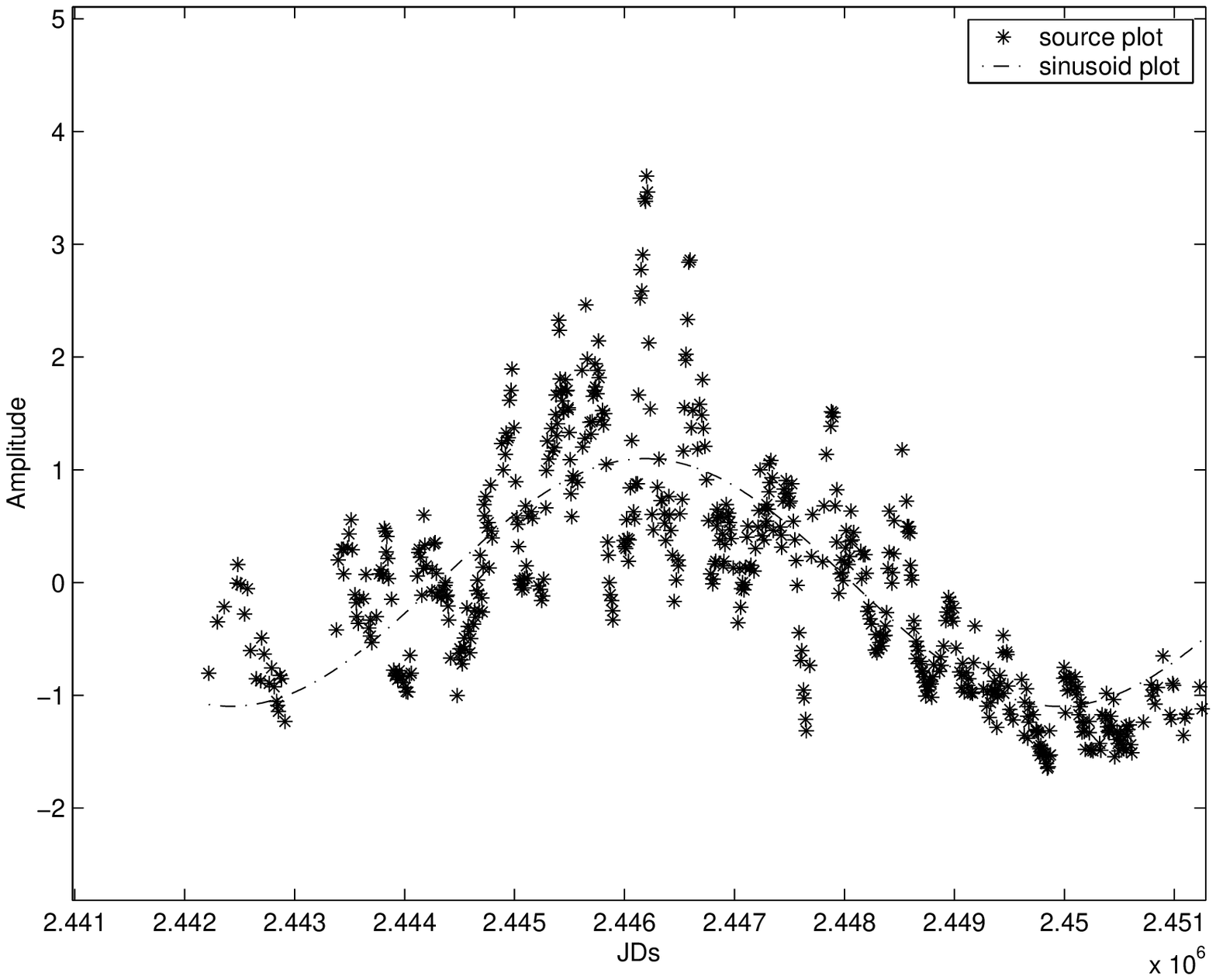}c)
\includegraphics[height=4cm, width=6cm]{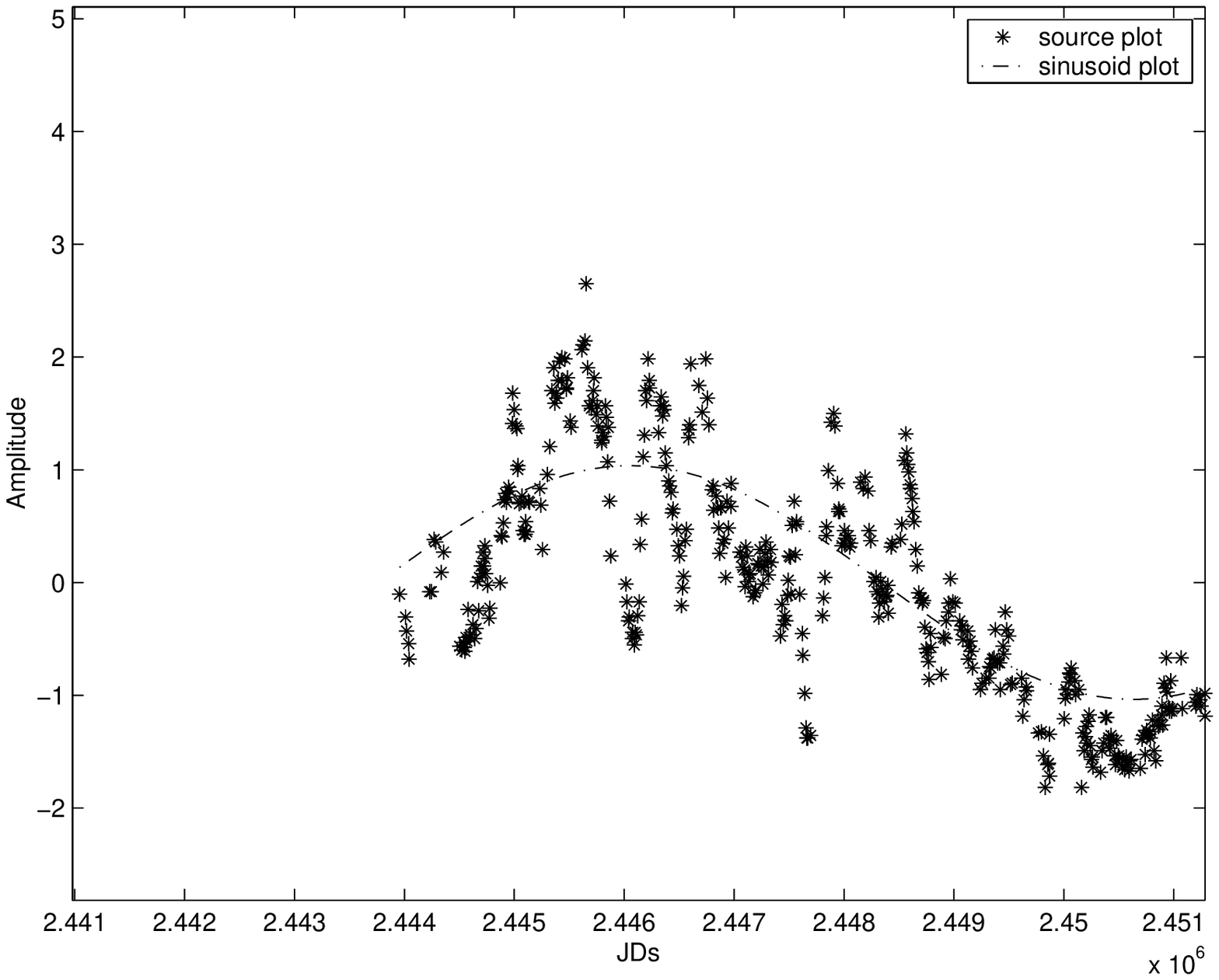}d)
\caption{OJ287. Panels a), b), c), and d): light curves at $22$,
$37$, 14.5, and 4.8 GHz, respectively. The dot-dashed sinusoids
have periods of about 20.9 yr (22 GHz), 22.3 yr (37 GHz), 20.6 yr
(14.5 GHz), 25 yr (4.8 GHz).}\label{OJ287}
\end{center}
\end{figure*}

\section{The time series analysis}

In what follows, we shall assume $x$ to be a physical variable
measured at discrete times $t_ i$; $x(t_ i)$ can be written as the
sum of the signal $x_s$ and random errors $R$:
\begin{equation}
x_i = x(t_ i) = x_ s(t_ i) +R(t_ i) \ .
\end{equation}
The problem is how to estimate possible fundamental frequencies
which may be present in the signal $x_s (t_ i)$. In the case of
even sampling, many Fourier-like tools can be effectively used
(cf. Fernie 1979; Horne \& Baliunas 1986). These methods, however,
encounter problems when dealing with unevenly sampled data and
shortcuts like the use of interpolation to re-sample the data,
introduce a noise amplification which usually undermines the
subsequent use of Fourier based techniques which are notoriously
very sensitive to the noise level of the input data (cf. Kay 1988;
Marple 1987, but see also Horowitz 1974).

\subsection{The Periodogram ($P$) method}
The most commonly used tool for periodicity analysis of both
evenly and unevenly sampled signals is the so called Periodogram
(hereafter $P$), which is an estimator of the signal energy in the
frequency domain (Deeming 1975), and has been extensively applied
to the study of light curves of variable stars, both periodic and
semi-periodic.

In the case of even sampling, the Periodogram has a simple
statistic distribution (exponentially distributed for Gaussian
noise). This, however, is no longer true for unevenly sampled
data, thus making it difficult to control errors (Kay 1988; Marple
1987; Oppenheim \& Shafler 1965). Scargle (1982) and Lomb (1976)
introduced a modified form of Periodogram which takes explicitly
into account the effects of uneven sampling.

Let $f$ be the frequency and $\tau$ a shift variable, and let us
also suppose that we are dealing with an observation series formed
by $N$ points $x(n)$. Their mean and variance are given by:
\begin{equation}
\overline{x} = \frac{1}{N}\sum_{n=1}^{N} x(n)$ and $\sigma^2=
\frac{1}{N-1}\sum_{n=1}^{N} (x(n)-\overline{x})^2 \ .
\end{equation}
The normalized Lomb's $P^L$, i.e. the power spectrum as a function
of the angular frequency $\omega \equiv 2 \pi f >0$ is defined as:
\begin{eqnarray}
 P^{L}_{N}(\omega) &=&\frac{1}{2\sigma^2}\left[
\frac{\left[\sum_{n=0}^{N-1} (x(n) -\overline{x})\cos \omega(t_n
-\tau)\right]^2}{\sum_{n=0}^{N-1} \cos^2 \omega (t_n
-\tau)}\right] \nonumber \\
&+&\frac{1}{2\sigma^2}\left[ \frac{\left[\sum_{n=0}^{N-1} (x(n)
-\overline{x})\sin \omega(t_n -\tau)\right]^2}{\sum_{n=0}^{N-1}
\sin^2 \omega (t_n -\tau)}\right]
\end{eqnarray}
and $\tau$ is defined by the equation:
\begin{equation}
\tan (2\omega\tau ) = \frac{\sum_{n=0}^{N-1}\sin 2\omega
t_n}{\sum_{n=0}^{N-1}\cos 2\omega t_n} \ .
\end{equation}

\subsection{Autocorrelation matrix based analysis: the STIMA approach}
Recently, more effective techniques based on the analysis of the
signal autocorrelation matrix were introduced [for a detailed
exposition of the problem see Kay (1988) and Marple (1987)]. In a
previous paper (Tagliaferri et al. 1999, hereafter T99) some of us
introduced the so called STIMA algorithm, based on a particular
type of the MUlty SIgnal Classifier (MUSIC) (Oppennheim \& Schafer
1965) estimator specifically tailored to work with unevenly
sampled data and on a robust nonlinear PCA Neural Network used to
extract the principal components of the autocorrelation matrix of
the input sources. Without entering into details (which may be
found in T99) we now briefly summarize its main features.

Let us assume to have a signal with $p$ sinusoidal components
(narrow band). The $p$ sinusoids are modelled as a stationary
ergodic signal, and this is possible only if the phases are
assumed to be independent random variables uniformly distributed
in the interval $[0, 2\pi]$. To estimate the frequencies of these
components we exploit the properties of the signal autocorrelation
matrix (a.m.) which is the sum of the signal and the noise
matrices. The $p$ principal eigenvectors of the signal matrix
allow the estimate of frequencies; the $p$ principal eigenvectors
of the signal matrix are the same of the total matrix. In order to
extract the principal components we use a robust nonlinear PCA
Neural Network  and we apply a modified version of MUSIC that can
be applied directly on unevenly sampled data, to obtain the
periodicities (cf. T99 and references therein).

The STIMA process for periodicity analysis can be divided in the
following steps:
\begin{itemize}

\item \underline{Preprocessing}: we first calculate and subtract
the average pattern to obtain a zero mean process (Karhunen \&
Joutsensalo 1994, 1995).

\item \underline{Neural computing}: the fundamental learning
parameters are: i) the initial weight matrix; ii) the number of
neurons, which is the number of principal eigenvectors that we
need, and therefore is equal to twice the number of signal
periodicities (for real signals); iii) $\alpha$, the nonlinear
learning function parameter; iv) the learning rate $\mu$.
\end{itemize}

We then initialize the weight matrix $\mathbf{W}$ assigning the
classical small random values. Otherwise we can use the first
patterns of the signal as the columns of the matrix. Experimental
results show, however, that even though the latter technique
speeds up the convergence of our neural estimator ($n.e.$), it
cannot be used with anomalously shaped signals, such as
stratigraphic geological signals. We use a simple criterion to
decide whether the neural network has reached or not convergence:
we control when the Power Spectrum of a modified MUSIC estimator
is greater than zero. This modified MUSIC estimator can be written
as:
\begin{equation}
P_M =\frac{1}{M-\sum_{i=1}^{M}|\mathbf{e}_{f}^{H}\mathbf{w}(i)|^2}
\end{equation}

where $\mathbf{w}(i)$ is the $i-th$ weight vector after learning,
and $\mathbf{e}_f^H$ is the sinusoidal vector:
\begin{equation}
\mathbf{e}_f^H = \left[e_{f}^{t_0},
e_{f}^{t_1},...,e_{f}^{t_{L-1}}\right]^H
\end{equation}
and $t_0$, ..., $t_{L-1}$, are the epochs of the first $L$
samplings.

We then have the following general algorithm:

\noindent - STEP 1: initialize the weight vectors
$\mathbf{w}_0(i)$ $\forall i=1, ..., p$ with small random values,
or with orthonormalized signal patterns. Then initialize the
learning threshold $\epsilon$ and the learning rate $\mu$. Then
reset pattern counter $k = 0$.\\
\noindent - STEP 2: input the $k-th$ pattern $$\mathbf{x}_k =
\left[x(k), ..., x(k+N+1) \right]$$ where $N$ is the number of
input components.\\
\noindent - STEP 3: calculate the output for each neuron
$y(j)=\mathbf{w}^T(j) \mathbf{x}_i$, $\forall i = 1, ..., p$.\\
\noindent - STEP 4: $\forall i = 1, ..., p$ modify the weights
$$\mathbf{w}_{k+1}(i) = \mathbf{w}_{k}(i) + \mu_{k}g(y_k(i))
\mathbf{e}_k(i)$$.\\
\noindent - STEP 5: convergence test. If
$$
P_M \frac{1}{M-\sum_{i=1}^{M}|\mathbf{e}_{f}^{H}\mathbf{w}(i)|^2}
> 0$$, then GO TO STEP 7.\\
\noindent - STEP 6: $k=k+1$. GO TO STEP 2.\\
\noindent - STEP 7. End.

The frequency estimator MUSIC takes as input the weight matrix
columns after the learning phase. The estimated signal frequencies
are obtained as the peak locations of the function previously
introduced. When $f$ is the frequency of the $i-th$ sinusoidal
component, $f=f_i$ , we have $P_M \rightarrow \infty$. In
practice, we have a peak near and in correspondence to the
component frequency. Estimates are therefore related to the
highest peaks.

We note that MUSIC works on individual points, not on the basis of
time intervals; it therefore does not need any kind of
interpolation to deal with gaps in the data. Also, tests carried
out by the Authors of this method (Tagliaferri et al. 1999) have
shown that low-frequency drifts of the baseline flux of sources do
not affect the derived periodicities.

In the following discussion we shall make use of both STIMA and
the Lomb Periodogram.

\section{Analysis of the Mets\"ahovi radio light curves}

We have analyzed the 22 and 37 GHz, both daily and weekly
averaged, light curves of blazars monitored with the Mets\"ahovi
radio telescope, updated to the end of 2001.

In order to establish whether a periodicity is real or not we
adopted the following set of rules:
\begin{itemize}
\item the periodicity derived by both STIMA and Lomb's method must
be shorter than $1/2$ of the time coverage (in order to avoid
aliasing);

\item For a given object, periodicity must be found in
both the $22$ and the $37$ light curves and the periods must be
equal given a percentage of error (in the following we use a
percentage of $10\%$)
\end{itemize}

Of the 157 sources in the sample by Ter\"asranta et al. (1998), 80
could not be analyzed due to either a very short time coverage or
a too sparse distribution of the data points. Evidence of a
periodicity was found for 5 of the remaining 77 objects (see Figs.
\ref{A0224_1}--\ref{A2251_1}). The results for these, based on
daily averaged fluxes, are given in Table~\ref{perioMetsa_1},
where the left-hand side refers to 22 GHz data, the right-hand
side to 37 GHz. The periodicity, given in units of $10^3$ days, is
detected by both STIMA and Lomb's methods and both methods yield,
for each source,  values very close to each other; the maximum
difference is of $\simeq 8\%$. In the following, unless explicitly
noted, we shall always use the STIMA estimate, which is less
affected by spurious signals. If weekly, rather than daily,
estimates are used, to increase the S/N ratio, the estimated
periods are essentially unaffected (differences are at percent
levels), even though the significance of periodicity is somewhat
increased.

Table~\ref{perioUMRAO} summarizes the results of the analysis of
the UMRAO monitoring data for the five sources with detected
periodicity, while a compendium of STIMA periods at the five UMRAO
plus Mets\"ahovi frequencies is presented in
Table~\ref{periosummary}.

For the sources $0224+671$ (Fig.~\ref{A0224_1}) and $0945+408$
(Fig.~\ref{A094+540}) the number of data points is quite limited
and, correspondingly, the periods are ill-defined. In the case of
A$0224+671$ the analysis of UMRAO data yields significantly
different periods at 4.8 GHz compared to 8 and 14.5 GHz; all of
them are substantially higher (by roughly a factor of 2) than
those found from the Mets\"ahovi light curves. For $0945+408$
(Fig.~\ref{A094+540}) no periodicity was found at 4.8 and 8.0 GHz,
while an ill-defined period of $2690$ days (STIMA), about a factor
of 2 larger than found at the Mets\"ahovi frequencies, could be
present in the $14.5$ GHz signal. These sources are good
candidates for longer and more frequently sampled monitoring
campaigns, particularly at the Mets\"ahovi frequencies where the
time span of light curves is relatively short.

In the case of $1226+023$ (Fig.~\ref{A1226_1}), the analysis of
the UMRAO database yields well defined periods at all frequencies
($4.8$, $8$, and $14.5$ GHz). The periodic behavior is most
evident at $4.8$ and $8$ GHz, but the estimated period at 8 GHz is
lower by $\sim 20\%$ than at 4.8 GHz. At higher frequencies (14.5,
22 and 37 GHz) the periodic signal appears to be contaminated by
an additional impulsive component which becomes dominant at $14.5$
GHz. The estimated periods at all frequencies, except 8 GHz, are
consistent with each other to better than 10\%.

Periodicity is detected, by both the STIMA and the Lomb's methods,
in light curves of $2200+420$ at all frequencies
(Fig.~\ref{A2200_1}), and period estimates are all consistent with
each other{ , and close to the value obtained in the optical:
$7.8\pm 0.2\,$yr, i.e. $2849\pm 73\,$days (Marchenko et al. 1996;
Hagen-Thorn et al. 1997)}. The periodicity is again most evident
at $4.8$ and $8$ GHz, while at $22$ and $37$ GHz the residuals
show the presence of an additional impulsive signal. At $14.5$
GHz, the evidence of periodicity is marginal, although the
estimated period is very close to that found in the other bands.

As shown by Fig.~\ref{A2251_1}, $2251+158$ is probably our best
case. Strong and consistent periodic signals are detected at all
frequencies, most clearly at $22$, $4.8$ and $8.0$ GHz. At all
frequencies, however, an inspection of the the residuals obtained
by subtracting the first harmonic from the observed data,
evidences an additional, impulsive, source of luminosity
variations, over-imposed on the underlying periodic component.

It may look surprising that our list of sources with evidence for
periodicity does not include the most famous case, OJ287. The
reason is that the estimated period does not meet our criterion of
being shorter than $1/2$ of the time coverage. In fact, we detect
a possible modulation on a period of about 21--22 years at14.5,
22, and 37 GHz, and on a longer period (about 25 years) at 4.8 GHz
(see Fig.~\ref{OJ287}). These possible periodicities are about a
factor of 2 larger than found by analyses of optical data
(Sillanp\"a\"a et al. 1988, 1996; Pietil\"a et al. 1999). An
analysis of the residuals (after having subtracted the sinusoidal
modulation) detects a significant periodicity of $\sim 1.6\,$yr at
22, 37, and 14.5, in close agreement with that found by Aller et
al. (1992) using a Scargle periodogram and by Hughes et al. (1998)
using a wavelet transform. Data at 8 and 4.8 GHz indicate (in the
residuals) somewhat longer periods (1.7 and 2.05 yr,
respectively).

\begin{table*}
\centering {\caption{Mean values of the variability index.
\label{VI}}
\begin{tabular}{lrlllll} \hline
{Type}& \multicolumn{1}{c}{N} &\multicolumn{1}{c}{37
GHz}&\multicolumn{1}{c}{22 GHz}&\multicolumn{1}{c}{14.5
GHz}&\multicolumn{1}{c}{8 GHz}&\multicolumn{1}{c}{4.8 GHz} \\
\hline
LBL & 9 &0.74$\pm$0.27&0.73$\pm$0.26&0.69$\pm$0.19&0.68$\pm$0.18&0.59$\pm$0.16\\
FSRQ& 16&0.61$\pm$0.16&0.55$\pm$0.15&0.54$\pm$0.12&0.50$\pm$0.11&0.38$\pm$0.08\\
\hline
HPQ & 9 &0.65$\pm$0.23&0.60$\pm$0.22&0.56$\pm$0.17&0.51$\pm$0.16&0.38$\pm$0.12\\
LPQ & 7 &0.56$\pm$0.23&0.48$\pm$0.21&0.49$\pm$0.18&0.45$\pm$0.17&0.33$\pm$0.13\\
\hline
\end{tabular}}
\end{table*}

\begin{table*} \centering
\caption{KS test on the distributions of variability indices.
\label{tabSFVIKS}}
\begin{tabular}{lllllllllll} \hline
   & \multicolumn{2}{c}{37 GHz} & \multicolumn{2}{c}{22 GHz} & \multicolumn{2}{c}{14.5 GHz}
   & \multicolumn{2}{c}{8 GHz} & \multicolumn{2}{c}{4.8 GHz} \\
  & \multicolumn{1}{c}{d}  & prob & \multicolumn{1}{c}{d} & prob & \multicolumn{1}{c}{d}
   & prob & \multicolumn{1}{c}{d} & prob & \multicolumn{1}{c}{d}  & prob \\
  \hline
LBL vs FSRQ & 0.65 &$7.3\,10^{-3}$ & 0.67 & $5.7\,10^{-3}$ &  0.52
& $7.2\, 10^{-3}$
& 0.61 & $1.4\,10^{-2}$ & 0.67 & $1.8\,10^{-4}$\\
HPQ vs LPQ & 0.32  & 0.73 & 0.46 & 0.28 & 0.39 & 0.33 & 0.32 &
0.73 & 0.28 & 0.75 \\
\hline
\end{tabular}
\end{table*}

\begin{table*} \centering
{\caption{Mean values of structure function slopes.
\label{tabSFslope}}
\begin{tabular}{llllll} \hline
\multicolumn{1}{c}{Type}& \multicolumn{1}{c}{37 GHz} &
\multicolumn{1}{c}{22 GHz} & \multicolumn{1}{c}{14.5 GHz}
   & \multicolumn{1}{c}{8 GHz} & \multicolumn{1}{c}{4.8 GHz} \\
   \hline
all &$1.14\pm0.23$&$1.18\pm0.25$&$1.17\pm0.19$&$1.17\pm0.20$&$1.21\pm0.20$\\
LBL &0.98$\pm$0.37 &0.95$\pm$0.37&1.05$\pm$0.29&1.00$\pm$0.28&1.05$\pm$0.29\\
FSRQ&1.08$\pm$0.29 &1.16$\pm$0.31&1.15$\pm$0.25&1.19$\pm$0.26&1.23$\pm$0.27\\
 \hline
HPQ &1.05$\pm$0.40 &1.09$\pm$0.41&1.08$\pm$0.35&1.18$\pm$0.38&1.14$\pm$0.37\\
LPQ &0.96$\pm$0.43 &1.07$\pm$0.48&1.15$\pm$0.44&1.17$\pm$0.46&1.24$\pm$0.47\\
\hline
\end{tabular}}
\end{table*}

\begin{table*} \centering
{\caption{Mean values of the $\log$ of the structure function
timescales. \label{tabSFtimescale}}
\begin{tabular}{llllll} \hline
\multicolumn{1}{c}{Type}& \multicolumn{1}{c}{37 GHz} &
\multicolumn{1}{c}{22 GHz} & \multicolumn{1}{c}{14.5 GHz}
   & \multicolumn{1}{c}{8 GHz} & \multicolumn{1}{c}{4.8 GHz} \\
   \hline
all
&$0.04\pm0.07$&$0.10\pm0.08$&$0.26\pm0.09$&$0.27\pm0.09$&$0.40\pm10$\\
LBL &-0.03$\pm$0.07&0.20$\pm$0.16&0.20$\pm$0.13&0.04$\pm$0.12&0.29$\pm$0.16\\
FSRQ& 0.09$\pm$0.10&0.07$\pm$0.10&0.33$\pm$0.11&0.41$\pm$0.13&0.46$\pm$0.13\\
\hline
HPQ & 0.06$\pm$0.13&0.05$\pm$0.13&0.13$\pm$0.10&0.08$\pm$0.12&0.19$\pm$0.14\\
LPQ & 0.17$\pm$0.17&0.13$\pm$0.15&0.46$\pm$0.22&0.61$\pm$0.26&0.64$\pm$0.26\\
\hline
\end{tabular}}
\end{table*}

\begin{figure*}
\begin{center}
\includegraphics[height=15cm, width=16cm]{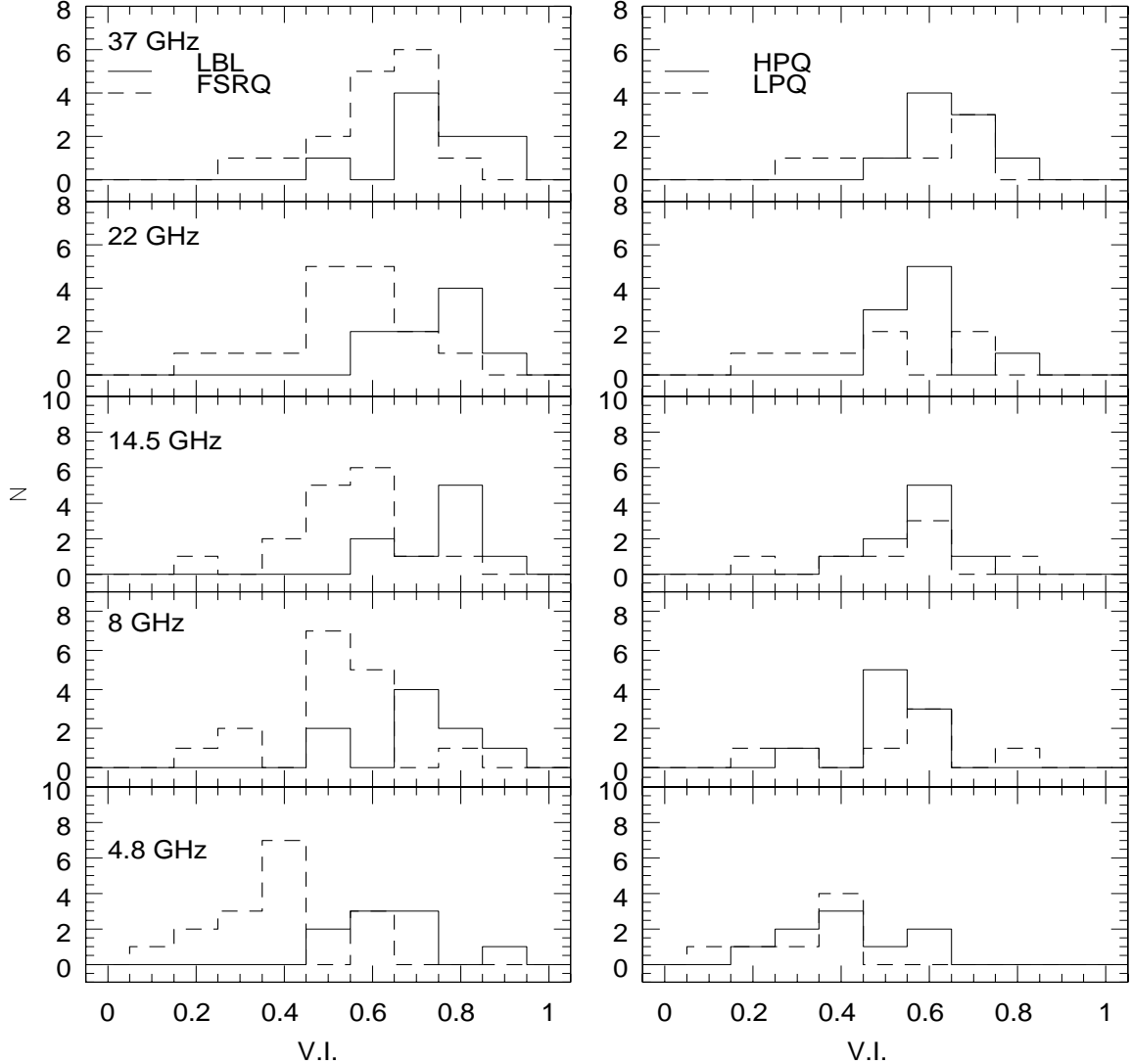}
\caption{Variability index distributions at different frequencies
for the U$\cap$M sample.}\label{VIplot}
\end{center}
\end{figure*}

\begin{figure*}
\begin{center}
\includegraphics[height=15cm, width=16cm]{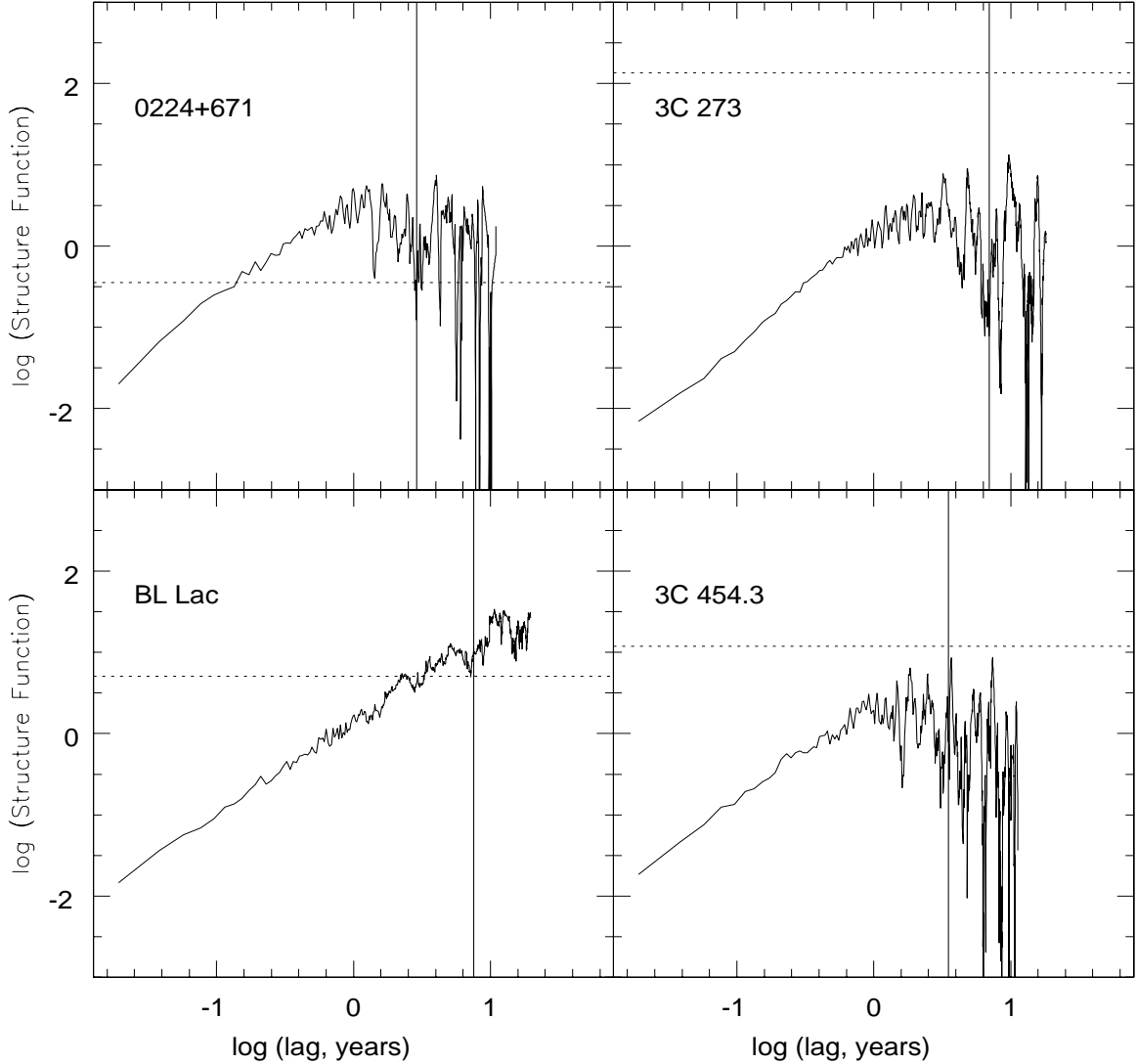}
\caption{Structure functions at 22 GHz of the 4 sources for which
we have the best evidences of periodicity. The values of the
periods are indicated by the vertical lines, while the dotted
horizontal lines show the variance of the process.}\label{SFper}
\end{center}
\end{figure*}

\begin{figure*}
\begin{center}
\includegraphics[height=15cm, width=16cm]{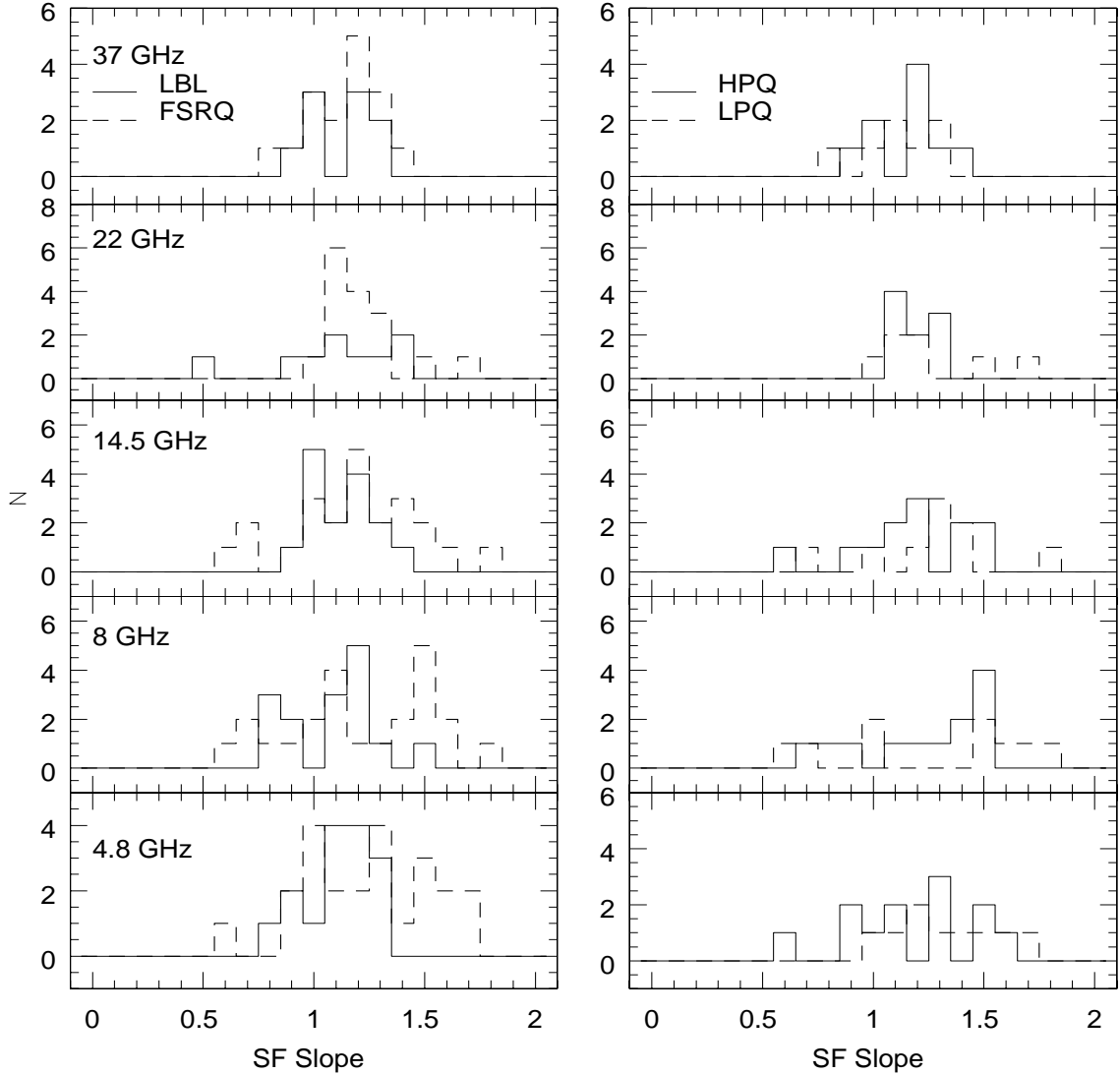}
\caption{Distributions of structure function slopes at different
frequencies. The results at frequencies $\le 14.5\,$GHz refer to
the full U sample, while at higher frequencies we are forcefully
limited to the 25 sources also monitored at Mets\"ahovi (U$\cap$M
sample).  }\label{SFslope}
\end{center}
\end{figure*}

\begin{figure*}
\begin{center}
\includegraphics[height=15cm, width=16cm]{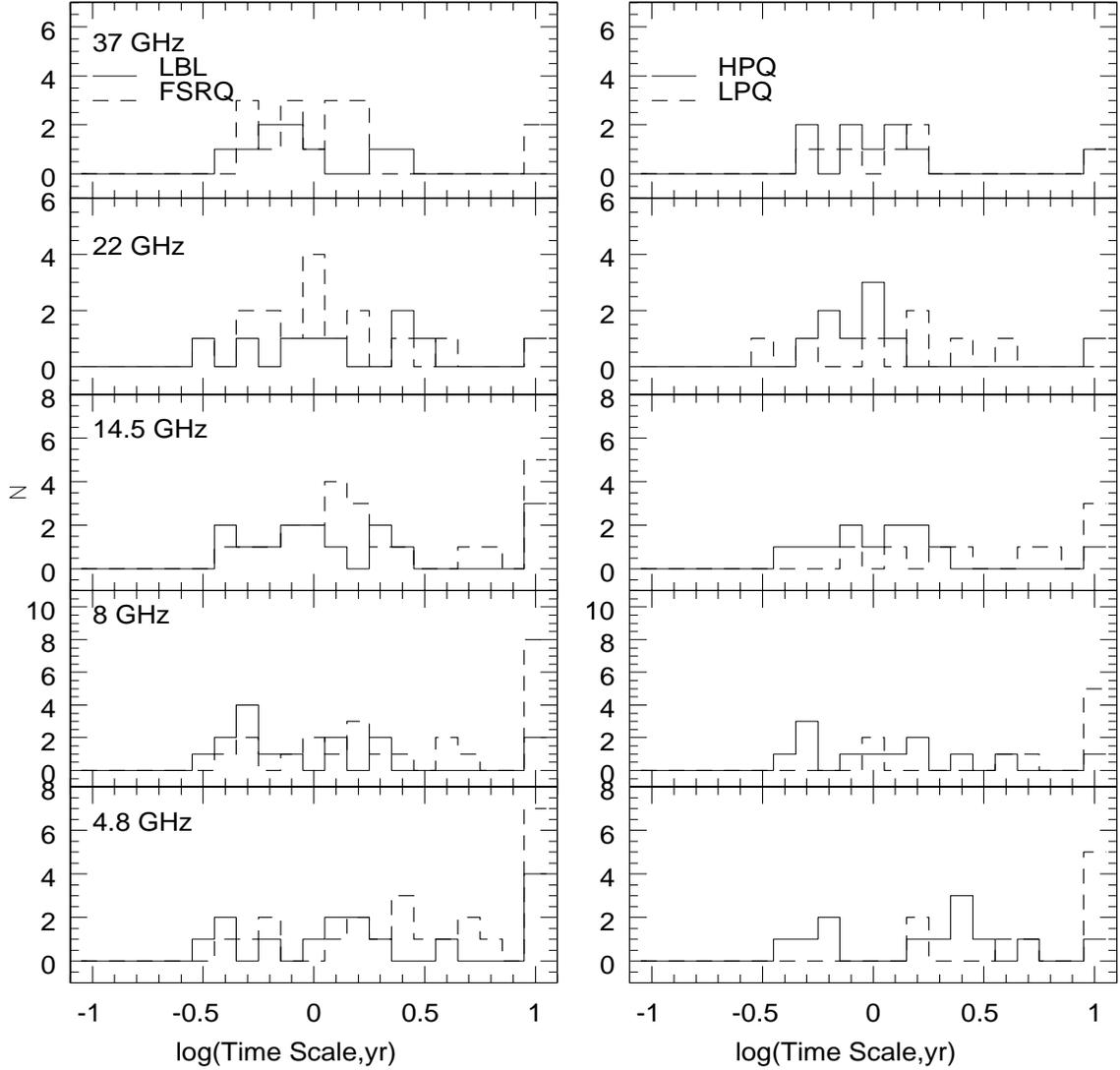}
\caption{Distributions of structure function timescales at
different frequencies. The samples used are the same as in
Fig.~\ref{SFslope}.} \label{SFtimescale}
\end{center}
\end{figure*}

\begin{figure*}
\begin{center}
\includegraphics[height=15cm, width=16cm]{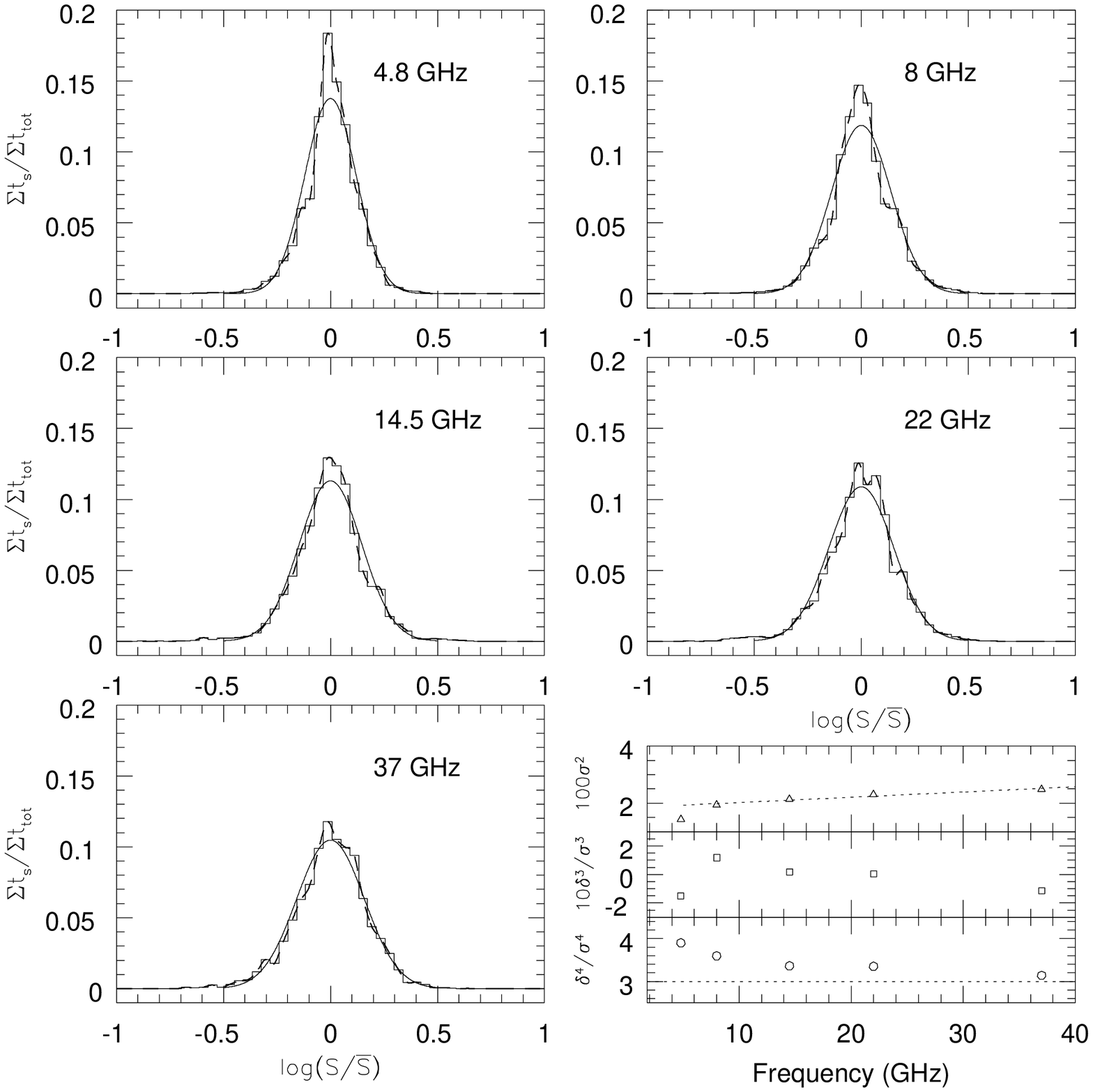}
\caption{Probability distribution functions of
$\log(S/\overline{S}$ at different frequencies and their Gaussian
fits. The panel in the lower right-hand corner shows, as a
function of frequency, the variance (multiplied by 100), the
skewness ($\times 10$), and the kurtosis. }\label{PDF}
\end{center}
\end{figure*}

\section{Phenomenology of variability properties}

The use of unbiased samples is essential for meaningful
statistical analyses. The UMRAO group have monitored and analyzed
the light curves of two complete samples, the Pearson-Readhead
sample (Aller et al. 2003) and a BL Lac sample (Aller et al.
1999).

The sample used in the present analysis is drawn from the complete
catalog of sources brighter than 1 Jy at 5 GHz (K\"uhr et al. {
1981}). We have have selected the K\"uhr sources with UMRAO
monitoring at 8 GHz for a total of $\ge 3000$ days, with gaps not
exceeding 200 days, all times being computed {\it in the source
frame} ($t_{\rm source}=t_{observer}/(1+z)$). Also, we have
confined ourselves to sources classified either as Low-energy peak
BL Lacs (LBLs, Padovani \& Giommi 1995) or Flat-Spectrum Radio
Quasars (FSRQ), thus excluding the lonely High-energy peak BL Lac
(HBL), namely Mkn~501, as well as the few sources classified as
radio-galaxies or Seyfert galaxies. Whenever available, we have
adopted the classification given by Donato et al. (2001),
supplemented with those by Ter\"{a}sranta  et al. (1998),
Ghisellini et al. (1993), and by Stickel et al. (1994).

The final sample comprises 39 sources (24 FSRQs and 15 LBLs),
listed in Table~\ref{Usample} (U sample). Of these sources, 25 (9
LBLs and 16 FSRQs) were also extensively monitored by the
Mets\"ahovi group (U$\cap$M sample). Of the 24 FSRQs, 21 are
classified as either HPQ (12) or LPQ (9); the acronyms stand for
high- or low-polarization quasars, respectively. Redshift
determinations are available for all sources. To characterize the
variability properties at the various frequencies we have computed
the variability index (V.I.), the structure function (Simonetti et
al. 1985; Hughes et al. 1992), and the distribution of intensity
variations. The variability index has been computed allowing for
measurement errors following Aller et al. (2003):
\begin{equation}
 V.I. =
\frac{(S_{max}-\sigma_{S_{max}})-
(S_{min}+\sigma_{S_{min}})}{(S_{max}-\sigma_{S_{max}})+
(S_{min}+\sigma_{S_{min}})} \ .
\end{equation}
Again following Aller et al. (2003) we have excluded anomalously
noisy data, i.e. data with $\sigma_S >\max(0.1\,\hbox{Jy},
0.03S)$.

\subsection{Variability index}

Table~\ref{VI} gives the mean values of the variability index
(V.I.) and their errors at 4.8, 8, 14.5, 22, and 37 GHz for the 25
blazars (9 LBLs and 16 FSRQs) in the U$\cap$M sample. Of the 16
FSRQs, 9 are HPQs and 7 are LPQs; we have also computed the V.I.
for each of these sub-classes. At all frequencies we get
systematically higher V.I.s for LBLs than for FSRQs (see also
Fig.~\ref{VIplot}), consistent with the results by Aller et al.
(1999). The Kolmogorov-Smirnov (KS) test shows that the
differences among the two populations are significant: the
probability that the two distributions come from the same parent
population is $\simeq 3.1\times 10^{-3}$, $\simeq 2.4\times
10^{-3}$, $\simeq 4.2\times 10^{-3}$, $\simeq 8.6\times 10^{-3}$,
and $9.1\times 10^{-5}$, at 37, 22, 14.5, 8, and 4.8 GHz
respectively (see Table~\ref{tabSFVIKS}).

The mean V.I. of FSRQ is higher at higher frequencies: it is
$\simeq 0.38$ at 4.8 GHz and becomes $\simeq 0.61$ at 37 GHz.
Again, the KS test confirm that the effect is statistically
significant: the probability that the distributions at 4.8 and 37
GHz come from the same parent population is $\simeq 4.9\times
10^{-5}$. The frequency dependence is less clear in the case of
LBLs, which have large V.I.s at all frequencies. Still, their mean
V.I. increases from $\simeq 0.59$ at 4.8 GHz to $\simeq 0.74$ at
$37\,$GHz; the significance of the difference  between the
distributions at the two frequencies is $\simeq 4.4\times
10^{-3}$. The frequency dependence of the V.I. is consistent with
the notion that, at lower and lower frequencies, an increasing
fraction of the observed emission is produced on larger and larger
scales and is therefore less variable on the timescales covered by
monitoring programs. At all frequencies the mean V.I. of
High-Polarization Quasars (HPQs) is slightly higher than that of
Low-Polarization Quasars (LPQs); the difference, however, has
always a low statistical significance.

\subsection{Structure function}

Another useful method for quantitatively investigating variability
properties is the so called structure function (S.F.) analysis
(Simonetti et al. 1985; Hughes et al. 1992). We have computed the
quantities commonly used to characterize the first order S.F.,
defined as $D(\tau)=\langle[S(t)-S(t+\tau)]^2\rangle$, namely its
slope ($d\log D/d\log\tau$) and the time-scale, i.e. the time lag
of the turnover of the S.F., which may measure the minimum time
scale of uncorrelated behavior or, in the case of flicker noise,
the minimum time scale in the distribution of response times
(Hughes et al. 1992).

In Fig.~\ref{SFper} we show the structure functions at 22 GHz for
the 4 sources for which we have the best evidence of periodicity.
The value of the derived period is indicated by the vertical line,
while the dotted horizontal line corresponds to the variance of
the process.

The distributions of slopes and of time-scales (in the source
frame) are shown in Figs.~\ref{SFslope} and \ref{SFtimescale},
respectively. Panels corresponding to frequencies of up to 14.5
GHz refer to the 39 sources in the U sample, those at higher
frequencies to the 25 sources in the U$\cap$M sample. Mean values
and dispersions are given in Tables~\ref{tabSFslope} and
\ref{tabSFtimescale}. Slopes and time-scales (or lower limits)
could be determined for essentially all sources. The exceptions
(sources with very irregular S.F.) are just one at 14.5 and 8 GHz,
and 3 at 4.8 GHz. For the sources (2 at 37 and 22 GHz, 10 at 14.5
GHz, 11 at 8 and 4.8 GHz) not showing a plateau we adopted the
time base of the data as a lower limit to the time-scale. We
further have a small number of sources with a plateau (1 at 37
GHz, 4 at 14.5 GHz, 1 at 8 GHz, and 2 at 4.8 GHz)  or a change of
slope (1 at 37 GHz, 2 at 22 GHz, 5 at 14.5 GHz, and 4 at 8 and at
4.8 GHz) at intermediate time lag, which may signal the transition
between two different processes; in these cases we have adopted
the longer time-scale. For sources with a change of slope we have
chosen the steeper value.

The average time-scale tend to be somewhat longer (although the
statistical significance of the difference between 37 and 4.8 GHz
is, for the full sample, only $1.2\times 10^{-2}$), and the
fraction of sources with time-scale exceeding the data time span
is larger at lower frequencies. The indication, reported by Hughes
et al. (1992), that LBLs have, on average, shorter timescales than
FSRQs is not statistically significant in the source frames (see
Table~\ref{tabSFtimescale}). There is a marginal indication
(significance of $2.4\times 10^{-2}$ based on the KS test at 8
GHz) of longer timescales for LPQs, compared to HPQs.

No statistically significant variations with frequency of the
slope distribution are detected. A visual inspection of
Fig.~\ref{SFslope} indicates that FSRQs tend to have steeper
slopes than LBLs. The average slope of LBLs is  $\simeq 1$, the
value corresponding to shot noise. In the the case of FSRQs, the
distributions are somewhat broader and extend to values $\geq
1.5$, particularly at lower frequencies, consistent with the
previous findings by Hughes et al. (1992). The KS test yields a
probability of $1.2\times 10^{-2}$ at 4.8 GHz and of $4.3\times
10^{-2}$ at 8 GHz that the distributions for the two populations
come from the same parent distribution. No statistically
significant differences are found at higher frequencies. The
distributions of LPQs and HPQs are consistent with the same parent
population at all frequencies .

Possible hints of a correlation of time-scales with slopes were
noted by Hughes et al. (1992). A Pearson's test however does not
detect any significant correlation among these quantities either
for the entire sample or for any sub-population.

\subsection{Distribution of intensity variations}

As shown by Fig.~\ref{PDF}, the distribution of
$\log(S/\overline{S})$, where $S$ is the instantaneous flux
density at a given frequency and $\overline{S}$ is its average
value, broadens with increasing frequency, consistent with the
notion the the variability amplitude increases with frequency
(Impey \& Neugebauer 1988). The frequency dependence of the
variance, $\sigma^2$, of the distribution is approximately linear.
A fit, shown as a dotted line in the upper part of the panel on
the lower right-hand corner of Fig.~\ref{PDF}, is given by:
\begin{equation}
\sigma^2 \simeq 1.846\times 10^{-4} \nu_{\rm GHz} + 0.01837 \ .
\label{sigma}
\end{equation}
The shape of the distribution also changes somewhat with
frequency. At high frequencies (22 and 37 GHz), the distribution
of intensity fluctuations approaches a log-normal distribution,
while at lower frequencies, the distribution of
$\log(S/\overline{S})$ is better described by a Laplace or by a
Cauchy distribution. The frequency dependencies of the skewness
and of the kurtosis are shown in the middle and on the bottom part
of the panel on the lower right-hand corner of Fig.~\ref{PDF}.

\begin{figure}
\begin{center}
\includegraphics[height=7cm, width=7cm]{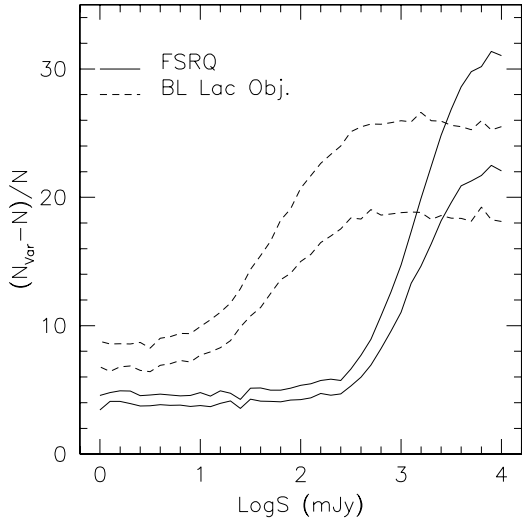}
\caption{Effects of variability on counts of FSRQs (solid lines)
and of LBLs (dashed lines). For each population, the upper curve
refers to 37 GHz, the lower one to 100 GHz. We have adopted a
Gaussian distribution for $\log(S/\overline{S}$ with variance
given by Eq.~(\ref{sigma}). Shown are the percentage differences
between the counts estimated allowing for variability and those
assuming that each source keeps at its average flux
$\overline{S}$.}\label{counts}
\end{center}
\end{figure}

Variability enhances the bright portion of the luminosity function
and of source counts. We have estimated its effect on the 37 and
100 GHz counts of FSRQs and of BL Lacs by convolving the
epoch-dependent luminosity functions given by Maraschi \& Rovetti
(1994) and Urry \& Padovani (1995), respectively, with the
distribution function of intensity fluctuation shown in the
corresponding panel of Fig.~\ref{PDF} at 37 GHz, and with a
Gaussian with variance given by Eq.~(\ref{sigma}) at 100 GHz. As
illustrated by Fig.~\ref{counts}, the enhancement of the counts is
of about 20--30\% at the brightest flux densities. It is thus very
likely that a substantial fraction of blazars picked up by the
high-frequency, shallow surveys carried out by the WMAP (Bennett
et al. 2003) and {\sc Planck}
satellites\footnote{astro.estec.esa.nl/SA-general/Projects/Planck/}
are in a flaring stage.

\section{Discussion and conclusions}

Combining the long-term monitoring databases of the University of
Michigan Radio Astronomy Observatory (UMRAO) and of the
Mets\"ahovi Radio Observatory it is possible to investigate the
radio variability properties of a rather rich sample of FSRQs and
LBLs over about a decade in frequency.

Our approach has followed different lines of investigation. First,
we have addressed the still controversial issue of the existence
of periodicities in the radio light curves. In addition to the
well known Periodogram method, modified to take into account the
effect of uneven sampling, we have applied a more effective
technique, never used before in this context, based on the
analysis of the signal autocorrelation matrix. We have analyzed
the Mets\"ahovi light curves looking for periods shorter than 50\%
of the maximum admissible period (to avoid aliasing) and
detectable (with the same value within the errors) at both 22 and
37 GHz. We have found evidences, confirmed by both techniques, for
periodicities in the light curves of 5 ($0224+671$, $0945+408$
$1226+023$, $2200+420$, and $2251+158$) out of the 77 Mets\"ahovi
sources having sufficient time coverage for a meaningful analysis
to be carried out. For the last three of these sources, consistent
periods are found also at the three UMRAO frequencies, while in
the case of $0224+671$ and $0945+408$ (which were monitored for a
relatively short time at Mets\"ahovi) UMRAO data give hints of
ill-defined periods a factor of 2 larger than at the Mets\"ahovi
frequencies.

The Lomb periodogram method allows us to test quantitatively the
significance of the detected periodicities. As shown by Scargle
(1982), the false-alarm probability, i.e. the probability that, if
we scan $M$ independent frequencies, none has spectral power
normalized to the variance of the data larger than $z$, under the
null hypothesis that the data are independent random Gaussian
values, is:
\begin{equation}
P(>z) = 1-(1-e^{-z})^M \ .
\end{equation}
An accurate evaluation of $P(>z)$ is complicated by the difficulty
of estimating $M$ (Press et al. 1996, p. 570). However, for any
reasonable choice of $M$, $P(>z)$ turns out to be extremely small
(and therefore the significance of the peak in the power spectrum
corresponding to the detected period is very high) for $1226+023$,
$2200+420$, and $2251+158$ ($z$ amounts to several tens at all
frequencies). On the other hand, the false alarms probability may
be quite significant for $0224+671$ and $0945+408$. As a further
test, we have applied the Kolmogorov-Smirnov test to examine the
consistency of the data distribution with a Gaussian random
process. This possibility is ruled out for $1226+023$ and
$2251+158$, but not for the other sources.

We have also investigated the variability index, the structure
function, and the distribution of intensity variations of the most
extensively monitored sources. We considered UMRAO sources
classified either as Low-energy peak BL Lacs or Flat-Spectrum
Radio Quasars, monitored at 8 GHz for a total of $\ge 3000$ days,
with gaps not exceeding 200 days, all times being computed {\it in
the source frame}. The sample comprises 39 sources (24 FSRQs and
15 LBLs), 25 of which (9 LBLs and 16 FSRQs) were also extensively
monitored by the Mets\"ahovi group. We have found a statistically
significant difference in the distribution of the variability
index for LBLs compared to FSRQs, in the sense that the former are
more variable. { This difference may help shedding light on the
relationship between the two blazar sub-classes and is consistent
with LBLs having a higher ratio of beamed to unbeamed emission
than FSRQs (Fan 2003).}

For both populations the variability index steadily increases with
increasing frequency. The distribution of intensity variations
also broadens with increasing frequency, and approaches a
log-normal shape at the highest frequencies.  Variability enhances
by 20--30\%, at bright flux densities, the high frequency counts
of extragalactic radio-sources, such as those which will be
carried out by the WMAP and {\sc Planck} satellites.

Since the sample is not complete, one should worry about possible
selection biases. To test this possibility we have extracted from
it a sub-sample 80\% complete to a flux limit of 1.2 Jy over the
three areas $16^h \le \alpha \le 20^h$, $53^{\deg} \le \delta \le
75^{\deg}$; $21^h \le \alpha \le 23^h$, $-8^{\deg} \le \delta \le
45^{\deg}$; $8^h \le \alpha \le 14^h$, $30^{\deg} \le \delta \le
45^{\deg}$. Such sub-sample comprises only 16 objects (6 LBLs and
10 FSRQs) and is thus too small for our purposes. On the other
hand, we did not detect significant differences in the mean
properties of sources in the sample of 39 compared to those in the
almost complete sub-sample. Also the trends with frequency and the
differences among sub-populations are still present. We are
therefore confident that, although the sample is not complete, it
is essentially unbiased.

\begin{acknowledgements}
      This research was partly supported by the Italian Space
      Agency (ASI) and by the Italian MIUR through a COFIN grant.
      The research by H.D. Aller and M.F. Aller has been supported
      in part by a series of NSF grants, including AST-9900723.
      The operation of UMRAO is
      supported by funds from the University of Michigan
      Department of Astronomy. { We are grateful to the
      referee, Dr. J.H. Fan, for helpful comments.}
\end{acknowledgements}

\end{document}